\documentclass[reprint, amsmath, amssymb, aps]{revtex4-1}
\usepackage{xcolor}
\usepackage{graphicx}
\usepackage{dcolumn}
\usepackage{bm}
\newcommand{\beq}{\begin{equation}}
\newcommand{\eneq}{\end{equation}}
\begin{document}

\title{Lindblad master equation approach to the topological phase transition in the disordered Su-Schrieffer-Heeger model}
 
\author{Andrea Nava$^{(1,2)}$, Gabriele Campagnano$^{(3)}$, Pasquale Sodano$^{(4)}$, and Domenico Giuliano$^{(1,2)}$ }
\affiliation{
$^{(1)}$Dipartimento di Fisica, Universit\`a della Calabria Arcavacata di 
Rende I-87036, Cosenza, Italy \\
$^{(2)}$I.N.F.N., Gruppo collegato di Cosenza, 
Arcavacata di Rende I-87036, Cosenza, Italy \\
$^{(3)}$CNR-SPIN, Monte S. Angelo-Via Cintia, I-80126, Napoli, Italy \\
$^{(4)}$I.N.F.N., Sezione di Perugia, Via A. Pascoli, I-06123, Perugia, Italy}

\begin{abstract}
 We  use the  Lindblad equation method to investigate the onset of a mobility edge  and the 
topological phase transition in the  disordered SSH chain connected to 
two external baths in the large bias limit. From the scaling properties of the nonequilibrium  stationary current flowing across the system,
 we recover  the localization/delocalization  in the disordered chain.
To probe the topological  phase transition in the presence of disorder, we  use  the  even-odd differential occupancy
as a mean to discriminate topologically trivial from topologically nontrival phases in the out-of-equilibrium system. 
 Eventually, we argue how to generalize  our method to other systems undergoing a topological phase transition in
the presence of disorder.  
\end{abstract}
\date{\today}
\maketitle

\section{Introduction}
\label{intro}

A common way to discriminate between different phases of matter is by
looking at their symmetries. Ordered phases  are typically characterized 
by a lower level  of symmetry than disordered ones. In many cases, this   
allows for introducing an order parameter as the average value of a local observable, 
which is different from  (equal to) 0 in an ordered (disordered) phase.  
Such an approach does not apply to topological phase transitions. 

The concept of topological phase has been originally  introduced within 
 the theoretical investigation of the Quantum Hall effect \cite{Thouless1982}, 
 as  a phase not characterized by  any symmetry breaking mechanism, but rather by fundamental, ``topological'' properties 
that are insensitive to smooth changes in the quantum state of the system 
\cite{Kane2005,Kane2005b,Bernevig2006}.

Topological phases
 are also of remarkable  interest for practical applications, as all of them 
are characterized by   nontrivial edge or surface states, which, being  protected by 
the topology of the phase itself,  have potential applications ranging from spintronics to 
topological quantum computation \cite{Nayak2008}. It is, therefore, 
of the utmost importance to have a criterion to  distinguish topologically nontrivial phases
from topologically trivial ones. Despite that, it is in general not simple to select 
a physical quantity playing the role of an order parameter at a topological phase
transition.   
 
 For lattice models that are translationally invariant by finite, lattice step translations, 
the key quantity  sensible to topology is the Chern number, defined  as the integral 
of the Berry curvature 
over the Brillouin zone of the system \cite{Berry1984,Kane2013}. By definition the Chern number is a collective 
property of the state of the system that must 
be an integer: it is naturally quantized and cannot be altered by a smooth deformation of 
  the state  which does not alter its  global topological properties. It is 
either 0, or different from zero, depending on whether the system  is topologically 
trivial, or not. In one-dimensional systems, the Chern number is proportional to the 
charge polarization, which measures the end charge
of the system, $Q_{\rm end}$ \cite{Zak1989,Resta1994,Kane2013,Lee2022}.  $Q_{\rm end}$ is different from zero provided 
the system hosts nontrivial edge state and, therefore, it is sensible to whether the system 
is in a topologically trivial or nontrivial state \cite{Kane2013}. 

When computed in a one-dimensional  system, it corresponds to the Zak phase \cite{Zak1989}, that is, to the integral of the Berry connection
over a closed path transversing the whole Brillouin zone. Importantly enough, nowadays technology 
already allows for a direct measurement of the Zak phase in cold atoms on optical lattices \cite{Atala2013}, 
with a good perspective of being soon able to perform similar measurements in solid-state devices. 

In real systems one has to face the unavoidable 
presence of the disorder, due to   impurities and/or defects. In particular, 
in one dimension, uncorrelated  disorder typically leads to full localization of the electronic 
wavefunctions and to the corresponding suppression of the current transport
\cite{Anderson1958,Abrahams1979}. In the specific context of topological materials, disorder 
can lead to reentrant topological phases. \cite{Pientka2013, Nava2017,Zuo2022}, with the corresponding onset of disorder-induced 
topological Anderson insulators \cite{Li2009}. Moreover, in the presence of correlated disorder, such as the random bond \cite{Flores1989},
 or the random dimer disorder \cite{Wu1992}, the interplay of disorder and topology can give rise to novel phases with 
remarkable properties, which 
are presently the subject of an intense research activity, on the theoretical \cite{Ostahie2021,Liu2022,Zuo2022,Wu2022}, 
as well as on the experimental  side, in solid-state systems, as well as in
 cold atom systems \cite{Meier2016,Gadway2015,Schaff2010}.

Over-all, the largest part of the theoretical, as well as of the experimental research has been 
focused onto the Su-Schrieffer-Heeger (SSH) model for polyacetilene   \cite{Heeger1988}. 
This is motivated on one hand by the fact that, despite its apparent simplicity, the theoretical 
model of SSH is able to catch the relevant physics arising from the interplay between topology 
and disorder, on the other hand by the relative ease to experimentally realize the model in 
setups like e.g. GaAs-AlGaAs superlattices \cite{Bellani1999}, 1D waveguide array fabricated in fused silica \cite{Naether2013}, 
or one-dimensional two-component ultracold atomic mixture in an optical lattice \cite{Schaff2010}.

Investigating the phase diagram of the disordered SSH model  
  presents two relevant kinds of difficulties. First of all, even in the absence of 
disorder, the SSH model is insulating. This implies that, when investigating 
the disorder-induced delocalization-localization phase transition, one cannot rely on the suppression of 
linear electric conductance as a function of the system size, since charge transport is already
suppressed by the gap,  even in the clean limit. Moreover, disorder breaks the lattice translational invariance, 
which invalidates the calculation of most of the quantities usually employed to distinguish the topologically nontrivial phase
from the topologically trivial one.

In this paper we apply the Lindblad master equation (LE) formalism \cite{Lindblad1976} to investigate both 
the localization-delocalization transition and the topological transition  
in a disordered open SSH chain connected to two external reservoirs (the ``baths'') in the large bias limit between the baths.
LE approach   allows to model, on very general grounds,
 the Markovian dynamics of an open quantum system connected to one or several baths. 
 In the recent years it has played a crucial role in different contexts, such as  ultracold atoms \cite{ Braaten2017,Lee2019}, 
 condensed matter systems \cite{Lieu2020, Budich2015,Nava2019,Nava2022_2,Nava2022_3}, quantum biology and quantum chemistry 
 \cite{Plenio2008, Manzano2013, Pino2015, Nava2022},
  with the possibility to implement quantum algorithms and experimentally realize Markovian dynamics 
  \cite{Nurdin2009,Verstraete2009,Diehl2008,Schlimgen2022}.
Furthermore, the Lindblad approach has recently  been used to induce topological phase transitions, in one- and two-dimensional systems, leading to a new universality class of Anderson transitions \cite{Diehl2011,Goldstein2019,Shavit2020,Beck2021}.
In the context of quasi-one-dimensional systems, the LE has been used to investigate both relaxation 
dynamics toward a thermal state, in terms of localized, or extended, bulk Lindblad operators
 \cite{Tarantelli2022,DiMeglio2020, Leeuw2021,Dangel2018,Artiaco2021}, as well as the
  non-equilibrium steady states (NESS)s that emerge when a system is
  placed in contact with two reservoirs at different temperatures or voltage bias/chemical potentials 
  \cite{Guimaraes2016,Tarantelli2021,Popkov2017,Iztok2013,Benenti2009b,Benenti2009,Nava2021,Maksimov2022}.
  
 In our specific case, the baths connected to the SSH chain  play the role of particle source and sink reservoirs and are modelled as 
 Lindblad local operators. Holding the baths at the large bias limit drives the system toward a NESS that 
  is characterized by a   steady-state value of the  charge current, $I_{\rm NESS}$. As we rigorously 
  prove analytically, and  evidence  numerically, $I_{\rm NESS}$ is finite in the absence of disorder,
  despite the system is gapped. Roughly speaking, this is due to the fact that  $I_{\rm NESS}$ is determined by states
  at all the energies, including, in particular, the conducting ones above the gap. Disorder-induced localization of the 
  states determines a suppression of $I_{\rm NESS}$ as a consequence of the corresponding localization of 
  single-particle wavefunctions. This eventually makes $I_{\rm NESS} \to 0$ as the chain length $L \to \infty$
  whenever all the single-particle states are localized. Following this observation,  we identify 
   the localized/delocalized phases of our system  according to whether $I_{\rm NESS}$ is suppressed/keeps finite
    as $L \to \infty$.   
   
 We therefore evidence how, once, in the large bias limit,  the chain is taken to a specific NESS, the even-odd differential occupancy (EOD) 
can be efficiently used to probe the nontrivial topological properties of the NESS itself. Specifically, by combining 
analytical and numerical methods, we show how a value of the EOD of about $\pm 1$ is directly related to 
the existence of the ingap states that characterize the topological phase, both in the clean limit and in 
the presence of (bond or dimer) disorder. 
We therefore conclude that our LE approach to the disordered, open SSH chain allows to encompass 
at the same time both the localization/delocalization transition, as well as the topological properties of the system, which 
eventually allows us to map out the corresponding  phase diagram with respect to both physical properties.
Eventually, after proving the effectiveness of our method, we argue how it can be potentially extended to 
disordered physical systems with nontrivial topological properties other than the SSH chain.

The paper is organized as follows:

\begin{itemize}
\item In section \ref{sec:model}          we introduce the model Hamiltonian for the SSH chain in the clean limit and 
analyze the two different types of disorder we consider here. Moreover, we present the LE approach and how we
apply it to our system.

\item In section \ref{transport_NESS}          we derive the $I_{\rm NESS}$ both in the clean limit and in the presence of disorder. 
In particular, we show how to map out the localization/delocalization phase transition from the behavior of 
$I_{\rm NESS}$ as $L$ gets large.

\item In section \ref{eod_0}          we introduce the EOD as a collective property of the NESS and show how to use it 
to probe the topologically nontrivial/trivial nature of the out-of-equilibrium state. 

\item In section \ref{conclusions}          we  summarize our results and provide possible further developments of our work.

\item In appendix \ref{derk}  we outline the derivation of the Eq.(\ref{eq:HF-master}) of the main text, 
while in appendix \ref{solclean}  we review the solution of the  SSH model over a finite chain with open boundary conditions 
in the absence of disorder, as well as the calculation of the scattering amplitudes in the presence of an impurity in 
the chain.

\end{itemize}

\section{Model and methods}
\label{sec:model} 

In the following we introduce the Hamiltonian for the one-dimensional SSH 
model   \cite{Heeger1988} and   present the Lindblad equation approach, which we 
employ to drive the system toward a NESS and to measure the physical quantities we use to 
characterize the phase diagram of the model, in the clean limit as well as in the presence of disorder. 

\subsection{Model Hamiltonian}
\label{modham}

Over a one-dimensional, $L$-site lattice, the Hamiltonian of the   SSH  model is given by  \cite{Heeger1988}

\beq
H_{\rm SSH} = - \sum_{ j = 1}^{L-1}                J_{j,j+1} \{ c_j^\dagger  c_{j+1}                + c_{j+1}^\dagger  c_j \}                - \sum_{ j = 1}^L \mu_j c_j^\dagger c_j 
\:\:\:\: . 
\label{model.1}
\eneq
\noindent
In Eq.(\ref{model.1}), we respectively denote with $c_j , c_j^\dagger$ the annihilation and the creation operator for a single spinless
fermion at site $j$ of the lattice, obeying the standard anticommutation algebra $\{ c_j , c_{j'}^\dagger \} = \delta_{j,j'}$. 
With $J_{j,j+1}$ and $\mu_j$ we respectively denote the single-fermion hopping amplitude between site $j$ and site $j+1$ and the 
chemical potential at site $j$. In the absence of impurities, 
we set $\mu_j = 0$ throughout the whole lattice, while for the $J_{j,j+1}$, consistently with the onset of dimerization in real polyacetilene 
in the presence of Peierls instability \cite{Heeger1988}, we set 

\beq
J_{j,j+1} = \Biggl\{ \begin{array}{l} J_o \:\: , \: {\rm for } \: j \: {\rm odd} \\
 J_e \:\: , \: {\rm for } \: j \: {\rm even} 
 \end{array}
 \:\:\:\: . 
 \label{model.2}
 \eneq
 \noindent
 Assuming periodic boundary conditions  and for $\mu_j = 0$  $\forall j$,  the SSH model undergoes a topological phase transition  
  as a function of $J_o/J_e$.  This corresponds to the (quantized)  Berry phase associated to the occupied Bloch wavefunctions to 
  switch from 0 to a finite value ($=\pi$ mod $2 \pi$)  \cite{Zak1989,Resta1994}. 
  
  The physical consequences of the topological phase transition are 
 evidenced  through the properties of the single-particle spectrum in  the chain with open 
 boundary conditions -- $H_{\rm SSH}$ in    Eq.(\ref{model.1}) --, as a function of $L$. Specifically, 
  due the Lieb theorem for dimerized systems \cite{lieb,mielke}, if $L$ is odd the system has at least a zero 
  energy state at any values of $J_o/J_e$. In this case $H_{\rm SSH}$ it is symmetric under $J_o\leftrightarrow J_e$ and no topological transition appears  at any value of $J_o/J_e$. 
At variance, for $L$ even,  two splitted in-gap states emerge in the single-particle spectrum   for $J_o<J_e$ with opposite values of the energy.  
 These two states correspond to a ``polarization charge'' $\pm e/2$ at the boundary of the chain \cite{Heeger1988}: they  become 
 strictly degenerate only in the thermodynamic limit, $L \to \infty$, in which case it is possible to linearly combine the 
corresponding wavefunctions into two orthogonal ones localized in real space near by either boundary of the chain. Due to this, 
from now on we shall  assume $L$ even throughout all our paper.

In the following we investigate the effects of 
two different kinds of disorder: the ``bond'' and the ``dimer'' disorder. To induce bond disorder,  we assign to the ``odd'' bond coupling 
strength $J_o$ a value randomly selected between $J_o = 1$ and $J_o = 1-W$, with a binary 
probability distribution ${\cal P}_{\rm b}                 [ J_o ]$ given by 

\beq
  {\cal P}_{\rm b} [  J_o] =\sigma \delta \left( J_o -1 \right) +\left(1- \sigma \right) \delta \left( J_o -1 +W \right)
    \:\:\:\: ,
    \label{model.3}
\eneq
\noindent
so that,  at each odd bond of the chain we may have single electron hopping $J_o$ either equal to 
 $1$, or to $1-W$,  with probability respectively given by $\sigma$ and $1-\sigma$. 
 Clearly, the $W=0$ limit corresponds to the clean case. 
To induce the dimer  disorder   we randomly assign  to the chemical potential at both sites of each elementary cell 
(that is, two consecutive odd and even sites)
either one  of two selected values, of which one is set at 0 \cite{Phillips1991,Zuo2022}.  The corresponding 
 probability distribution ${\cal P}_{\rm d} [ \mu ]$ is given by  

\beq
{\cal P}_{\rm d}           [            \mu_j  ] = \begin{cases}
\sigma\delta\left(\mu-0\right)+\left(1-\sigma\right)\delta\left(\mu-W\right) & j\ \mathrm{odd}\\
\mu_{j-1} & j\ \mathrm{even}
\end{cases}
\:\:\:\: , 
\label{model.4}
\eneq
\noindent
so that $\mu$ can be zero or $W$ with probability $\sigma$ and $1-\sigma$ respectively. The $W=0$ limit corresponds again to the clean case. 

      The two realizations of the disorder   in 
Eqs.(\ref{model.3}) and (\ref{model.4}) have  completely different physical consequences on 
the SSH chain, due to the different behavior of the corresponding potential under the action of 
the ``chiral operator''  ${\bf \Gamma} $ defined as

\beq
{\bf \Gamma}  = \sum_{j = 1}^\frac{L}{2}         \{ c_{2j-1}^\dagger c_{2j-1} - c_{2j}^\dagger c_{2j} \}
\;\;\;\; . 
\label{chiral.def}
\eneq

 In the clean limit one has   $ \{ {\bf \Gamma} , H_{\rm SSH} \}   = 0$, provided that the  chemical potential
is set to 0, which  leaves the spectrum of $H_{\rm SSH}$ invariant.  When disorder is added to the SSH chain, thus 
breaking lattice translational invariance, it is no more possible to define the Berry phase as the integral over the Brillouin zone of 
the Berry connection. However, if 
 the disordered  Hamiltonian, at any given realization of the disorder potential, still anticommutes with ${\bf \Gamma}$, 
 it is possible to  introduce a ``disorder-averaged real space winding number'' (DAWN), $\delta_\nu$, 
  which can be used to label topological phases in the presence of disorder.
  
   Specifically, in a one-dimensional system of length $L$, $\delta_\nu$ is computed 
  by ensemble averaging over a large enough number $N$ of independent  realizations of the 
  disorder, according to the formula  \cite{Mondragon2014,Liu2022}       
 
 \beq
 \delta_\nu = \frac{1}{N}   \sum_{s = 1}^N  \frac{1}{L}         \: {\rm Tr} \{         \Gamma Q_s [ Q_s , X] \}         
 \:\:\:\: . 
 \label{dawn.1}
 \eneq
 \noindent
 In Eq.(\ref{dawn.1})  $_s$ labels a single realization of the disorder, $Q_s =  \sum_n \{| \psi_{n.s}         \rangle \langle \psi_{n.s}         | - 
 {\bf \Gamma} | \psi_{n,s}         \rangle \langle \psi_{n.s}         | {\bf \Gamma}$, with $ \{         | \psi_{n.s} \rangle \}$ being a complete set 
 of eigenfunctions of the Hamiltonian at the realization of the disorder labeled with $_s$, and $X$ being the unit-cell 
 coordinate operator, $X = {\rm diag} \left( 1,1,2,2, \ldots , \frac{L}{2} , \frac{L}{2}         \right)$. 
 
 Due to the missing anticommutativity with ${\bf \Gamma}$, 
 the  DAWN cannot be defined in the presence of dimer disorder. 
This requires introducing alternative means to investigate   the combined effects of topology and disorder  \cite{Phillips1991,Zuo2022}. 
To bypass   this limitation, in the following we define and employ the  EOD.
We show that our method allows us to witness  the onset of topological phases in the presence of disorder
regardless of whether the Hamiltonian anticommutes with ${\bf \Gamma}$, or not. Moreover, in 
 the large bias limit, we show how our method  
can be  equally well applied to equilibrium, as well as to out-of-equilibrium, open systems. 

 To analyze  the disorder-induced localization effects on the electronic states
of the system, we study the stationary charge transport properties of the open chain connected to 
two external baths taken at large chemical potential bias.  Within the LE approach, we 
  derive the stationary density matrix $\rho$ describing the 
 NESS that asymptotically sets in the system. Computing
    the corresponding stationary current $I_{\rm NESS}$ flowing 
through the chain we will be able to spell out disorder-induced effects in the system through their effects in $I_{\rm NESS}$. 
 Motivated by this observation, we now present the main features of the LE  approach to the SSH model.    
 
\subsection{Lindblad equation}
\label{lineq}

The LE master equations consists of  a first order
differential equation for the time evolution of the system density matrix $\rho(t)$, 
given by 
\begin{equation}
\dot{\rho} ( t  )=
-i [H,\rho (t)  ]+\sum_{k} (L_{k}\rho ( t )  L_{k}^{\dagger}-\frac{1}{2} \{ L_{k}^{\dagger}L_{k},\rho (t)  \}  )
\:\:\:\: . 
\label{eq:lindbladeq}
\end{equation}
\noindent
The first term at the right-hand side of Eq.(\ref{eq:lindbladeq}), called the Liouvillian, describes the unitary evolution determined  by 
 the system Hamiltonian $H$, while 
the second term, the  Lindbladian, includes dissipation and decoherence on  the system dynamics, with the  
``jump'' operators $L_k$ that are determined by the coupling between the system and the baths. Throughout 
the following derivation, we assume $H = \sum_{i,j = 1}^L c_i^\dagger  {\cal H}_{i,j} ( t ) c_j$, with $c_j , c_j^\dagger$ being 
the single-fermion creation and annihilation operators at site $j$ and the Hamiltonian matrix elements 
$ {\cal H}_{i,j} ( t)$ which, in general, can depend on time $t$, as well.

In the following, we   consider  baths that locally inject (or extract) fermions to (from) the boundary sites of the chain, 
at  given and fixed rates. We describe the injecting and extracting baths at site $j=\left\{ 1,L\right\} $ in
terms of the Lindblad operators $L_{in,j} $ and $L_{out,j} $, given by 
\begin{eqnarray}
L_{in,j} & = & \sqrt{\Gamma_{j}}c_{j}^{\dagger} \nonumber \\
L_{out,j} & = & \sqrt{\gamma_{j}}c_{j} \:\:\:\: ,
\label{eq:deltastar-L}
\end{eqnarray}
\noindent
with $\Gamma_j$  and $\gamma_j$  being  the coupling strengths respectively determining the creation and the annihilation of a fermion at  site $j$.

We have a total of four, in principle independent, coupling 
strengths,  $\Gamma_1$, $\gamma_1$, $\Gamma_L$ and $\gamma_L$, that describe the coupling between the chain and the baths.
When recovering the above couplings from the microscopic theory, we see that they can be expressed in terms 
of the  Fermi distribution function  at the chemical potential of the reservoir, $f$ and of the reservoir  spectral density
 at the chemical potential of the reservoir, $g$. Specifically, we  obtain \cite{petruccione,zoller}
\begin{eqnarray}
\Gamma_{j} &=&g_{j}f_{j}\nonumber \\
\gamma_{j} &=& g_{j}(1-f_{j})
\;\;\;\; , 
\label{rates.1}
\end{eqnarray}
\noindent
with (labeling each reservoir with the index of the site it is connected to) $j=\left\{ 1,L\right\} $. 

 In this paper we focus onto the large bias regime, which corresponds to  setting 
$f_1=1$ and $f_L=0$.    In this  limit,   the  reservoir coupled to site $1$  acts as an electron ``source'', 
by only injecting electrons in the chain, and the reservoir   coupled to site $L$ 
acts as an electron ``drain'',  by only  absorbing electrons from the system. As a result, 
a steady-state current is induced, due to 
electrons that enter the chain at site 1 and  travel all the way down to site $L$, where 
they  exit the chain. To derive the current in the large bias limit, we 
first determine  $\rho ( t )$ by solving Eq.(\ref{eq:lindbladeq}). Then,   we compute
the (time dependent) expectation value of any observable $O$, $O(t)$ using 
\begin{equation}
O(t) =
{\rm Tr} [ O \rho\left(t\right) ] 
\:\:\:\: . 
\label{ave.1}
\end{equation}
\noindent
Taking into account Eq.(\ref{ave.1})  
we  employ Eq.(\ref{eq:lindbladeq}) to write down the LE directly 
for $O ( t ) $ obtaining 
\begin{eqnarray}
\frac{d}{dt} O ( t )   & = & {\rm Tr}
 ( O  \dot{\rho} (t )  )= i {\rm Tr} [[  H, O   ] \rho ( t )  ] 
\label{eq:observable-master equation}  \\ 
& + & \sum_{k} ( {\rm Tr} [  L_{k}^{\dagger} O  L_{k}\rho ( t ) ]  -
\frac{1}{2} {\rm Tr} [  \{ L_{k}^{\dagger}L_{k}, O 
 \} \rho ( t ) ]      )\nonumber
\:\:\:\: . 
\end{eqnarray}
\noindent
In the following we  focus on the average value of the occupation number at  a generic 
site $i$ of the system, $n_j ( t ) = {\rm Tr} [ n_j  \rho ( t )]$, as well as of the currents flowing from the
 reservoirs into the site $j=\left\{ 1,L\right\} $,  $I_{in,j} ( t ) $,  or from site $j=\left\{ 1,L\right\} $ to 
the reservoir, $I_{out,j} ( t )$.  These are given by 
\begin{eqnarray}
I_{in,j} ( t )  & = & \Gamma_{j} (1-  n_j ( t )) \nonumber \\
I_{out,j}( t )  & = & \gamma_{j}  n_j ( t ) 
\:\:\:\: , 
\label{curave.1}
\end{eqnarray}
\noindent
so that the net current exchanged at time $t$ 
between the reservoirs and the site $j$ is given by $I_{j} (t)  =  I_{in,j} ( t ) - I_{out,j} ( t )$.
In addition, we also need to derive the average value of the current flowing between two connected sites of the chain, say $j$ and $j\pm1$, $I_{j,j\pm1}$. This is given by 
\begin{equation}
I_{j,j\pm1} ( t ) = - i J_{j,j\pm1}   {\cal I}_{j,j\pm1} ( t )  + {\rm c.c.} 
\:\;\;\; , 
\label{curave.2}
\end{equation}
\noindent
with ${\cal I}_{j,j\pm1} ( t ) = {\rm Tr} [c_j^\dagger c_{j\pm1}  \rho ( t )]$ and  with ${\rm c.c.}$ denoting the complex conjugate.
 
The full set of  LEs for $n_j ( t )$,  $I_{in,j} (t)$,  $I_{out,j} (t)$, and $I_{j,j\pm 1} (t)$ allows us to 
recover the current across the chain when it is connected to external reservoirs. For a quadratic Hamiltonian,
 it is possible to write a closed set of equations for the  bilinear operators only that can   
be written in matrix form as
\begin{equation}
\dot{{\cal C}} ( t )  =i [  {\cal H}^{\top} ( t )   , {\cal C} ( t )  ]+ {\cal G}-
\frac{1}{2} \{  ( {\cal G}+ {\cal R} ), {\cal C} ( t )  \} 
\;\;\;\; , 
\label{eq:HF-master}
\end{equation}
with the bilinear expectation matrix elements  $[ {\cal C} ( t ) ]_{i,j}= {\rm Tr} [c_i^\dagger c_{j}  \rho ( t )]$, 
and the system-bath coupling matrix elements $ [ {\cal G}]_{i,j}=\delta_{i,j}\left(\delta_{j,1}\Gamma_{1}+\delta_{j,L}\Gamma_{L} \right)$ and $ [  {\cal R} ]_{i,j}=\delta_{i,j}\left(\delta_{j,1}\gamma_{1}+\delta_{j,L}\gamma_{L} \right)$.
The system evolves in time, asymptotically 
flowing to  the NESS, which is determined from the condition $\dot{ {\cal C}}(t)=0$  (details 
about the derivation   of Eq.(\ref{eq:HF-master})   are provided in appendix \ref{derk}). 

In the following we use  Eq.(\ref{eq:HF-master}) to describe the SSH model with different kinds 
of  correlated disorder.

\section{Current and state localization in the non-equilibrium steady-state}
\label{transport_NESS}

 The most effective way of probing the disorder-induced localization   in one-dimensional 
systems is through dc current transport measurements \cite{Giamarchi2003}. 
However, due to the presence of the dimerization gap, the SSH chain is insulating even in the absence of disorder.
In this case, as an alternative (to transport properties) mean to study the localization transition, it has been
proposed to look at the normal- and at the inverse-participation ratio. These two quantities can directly be computed from 
the wavefunctions for the single-electron states in the system.
By averaging over $N$ independent realizations of the disorder,  
they are respectively defined as  \cite{Liu2022}

\begin{eqnarray}
{\rm NPR} &=&\frac{1}{N} \sum_{s = 1}^N \frac{1}{L} \sum_n \left( L \sum_{j=1}^L | \langle j | \psi_{n,s} \rangle |^4 \right)^{-1} \nonumber \\
{\rm IPR} &=&  \frac{1}{N} \sum_{s = 1}^N  \frac{1}{L} \sum_n    \sum_{j=1}^L | \langle j | \psi_{n,s} \rangle |^4 
\:\:\:\: . 
\label{npipr.1}
\end{eqnarray}
\noindent
As  $L \to \infty$, either  ${\rm NPR}=  0$, corresponding to localization of 
all the states, or ${\rm IPR=0}$, corresponding to all the states being delocalized.

As we show in the following, driving the chain to the large-bias limit allows for using charge transport to probe the localization transition even for 
the insulating system. Indeed, in  the linear response regime the current is proportional to the 
zero-energy transmission 
coefficient ${\cal T}$ across the chain and is therefore exponentially suppressed with $L$ as $e^{ - \frac{L}{\xi}}$, 
with $\xi = 1 / \tanh^{-1}          \left[          \frac{ | J_e - J_o |}{ J_e + J_o }          \right]$. Instead,
once the system is driven toward  the NESS  corresponding   to the ``optimal working point'' 
by tuning   the chain-bath coupling strength  
  \cite{Benenti2009,Nava2021},  $I_{\rm NESS}$ keeps 
finite as $L \to \infty$ even if the system is gapped.  

On turning on the disorder,  $I_{\rm NESS}$ is suppressed, due to the strong
localization effect of random disorder in one-dimensional systems \cite{Anderson1958,Abrahams1979}, thus 
signaling the onset of the delocalization/localization transition in the electronic states in the chain.    Moreover, 
as we show below, $I_{\rm NESS}$ is also sensible to the onset of the mobility edge that has been 
predicted to arise, under suitable conditions, in one-dimensional disordered systems in the presence of correlated disorder
and/or of ``structured'' impurities \cite{Dunlap1990,Phillips1991,Wu1992,Sedrakyan2004}  (see 
also \cite{Sedrakyan2004b} for an extension to a two-leg ladder and \cite{Sedrakyan2011}  for a proposal of a practical implementation 
within an N-leg ladder realized with an optical lattice):  the  mobility edge is indeed
``naturally'' revealed by $I_{\rm NESS}$ maintaining a finite value, even on increasing $L$ toward the scaling limit. 
Thus, we conclude that  measuring  $I_{\rm NESS}$ is enough to determine  the localization/delocalization  
transition in our system.  At variance,  to discriminate between the topologically trivial and the topologically nontrivial phase,
below we introduce  the EOD as a suitable tool to infer   the topological 
properties of the NESS.
 
 To  effectively ascertain the physical 
behavior of our system and how it is affected, in the following we employ 
 analytical and numerical methods, by complementing    them one another, 
 whenever possible.

\subsection{$I_{\rm NESS}$   in the clean limit}
\label{iness_clean}

 In the absence of disorder,  we compute $I_{\rm NESS}$  by 
employing  the   approach of \cite{Guimaraes2016}. 
The basic idea is to think of a three-region system: region $A$ and $C$, respectively 
connected to the left-hand and to the right-hand bath, and the central region $B$,
whose sites are not directly connected to a bath and to eventually assume 
that $A$ and $C$ are  weakly coupled to $B$. Defining $g$ to be 
the small parameter controlling the coupling strength, so that  $0<g\ll 1$,
we compute the current to    leading order in  $g$.

 The constraint over $g$ does not affect the reliability  of the final result (at least 
qualitatively).  This is due to the absence of a ``bulk'' interaction 
in the SSH model, which makes the inter-site hopping between neighboring   a marginal 
boundary operator. Therefore, tuning its strength by continuously varying $g$ toward 
 the $g \to 1$ limit (which corresponds to our specific case) does not qualitatively alter 
 the results we obtain in the small-$g$ limit  (note that this 
 does typically not happen in the case in which $g$ is the interaction strength in 
 front of a relevant, or of a marginally relevant, boundary interaction, such as,
 for instance, in the presence of Majorana modes at the endpoints of the chain \cite{Fidkowski2012,Affleck2013,Affleck2014}, 
 or in the two-impurity Kondo model in a spin chain \cite{Bayat2012}).

    \begin{figure}
 \center
\includegraphics*[width=0.8 \linewidth]{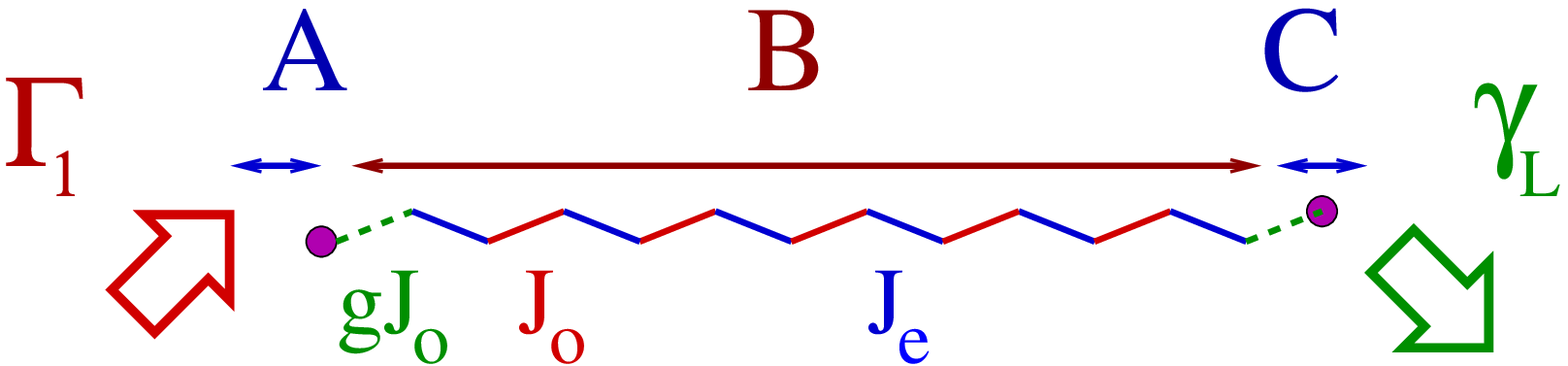}
\caption{  Sketch of the SSH chain connected to one-site Lindblad baths (drawn as full, purple dots) with 
weakened bonds $J_{1,2}                = J_{L-1,L} = g J_o$ (drawn in green).   In the figure we evidence 
the three regions of length A,B and C (see main text for details).   The 
red arrow evidences  the particle injection from the source to the chain (with coupling $\Gamma_1$) 
and the green arrow corresponds to the particle injection from the chain into the drain (with coupling
$\gamma_L$) when the system is taken to the large bias regime. } 
\label{sketch_single}
\end{figure}
\noindent
In the following , we realize the single-site 
 $A$ and $C$ regions   by  ``weakening'' the first and the last 
bond of $H_{\rm SSH}$ in Eq.(\ref{model.1}), according to 

\begin{eqnarray}
J_{1,2} &=& J_o \to g  J_o \nonumber \\
J_{L-1,L} &=& J_o \to g J_o 
\;\;\;\; ,
\label{ness.c.1}
\end{eqnarray}
\noindent

In Fig.\ref{sketch_single}          we provide a schematic drawing of our system, with the 
red arrow evidencing the particle injection from the source to the chain (with coupling $\Gamma_1$) 
and the green arrow corresponding to the particle injection from the chain into the drain (with coupling
$\gamma_L$) when the system is taken to the large bias regime.
The corresponding  Hamiltonian $  \hat{H}$ is determined by the sum of the Hamiltonian of 
 the two sites connected with the external baths, $ \hat{H}_A , \hat{H}_C$,  
   plus the terms coupling $A$ and $C$ sites to the middle chain, $ \hat{V}_{A ,B}$ and 
$ \hat{V}_{B, C}$, plus the Hamiltonian for the middle region $ \hat{H}_B$,  that is 

\beq
 \hat{H} = \hat{H}_A + \hat{H}_C + \hat{H}_B + \hat{V}_{A , B} + 
\hat{V}_{B, C}
\;\;\;\;.
\label{est.0}
\eneq
\noindent
Using the notation $\alpha , \alpha^\dagger$ and $\beta , \beta^\dagger$ to respectively denote the 
single fermion annihilation and creation operators over respectively the first and the last
site of the chain, we have

\begin{eqnarray}
 \hat{H}_A  &=& \epsilon_\alpha \alpha^\dagger \alpha \nonumber \\
 \hat{H}_C   &=& \epsilon_\beta \beta^\dagger \beta \nonumber \\
 \hat{H}_B  &=& - \sum_{ j = 1}^{L-3} J_{j , j+1}                \{ c_j^\dagger c_{j+1} + c_{j+1}^\dagger c_j \} \nonumber \\
 \hat{V}_{A , B}  &=& - g J_o \{\alpha^\dagger c_1 + c_1^\dagger \alpha \} \nonumber \\
 \hat{V}_{B,C } &=& - g J_o \{\beta^\dagger c_L + c_L^\dagger \beta \} 
\:\:\:\: , 
\label{est.1p}
\end{eqnarray}
\noindent
with  $\epsilon_\alpha , \epsilon_\beta$  eventually
sent to 0, consistently with \cite{Guimaraes2016} and with 
$J_{1,2} = J_e , J_{2,3} = J_o$, et cetera. Let us note that $ \hat{H}_B$ corresponds to the Hamiltonian of an SSH 
chain of $\hat{L}\equiv L-2$ sites where $J_e$ and $J_o$ are exchanged with each other.  

  From Eqs.(\ref{est.1p}) we  find that $I_{\rm NESS}$ is 
given by 

\beq
I_{\rm NESS} = - \frac{igJ_o}{2} \{ \langle \alpha^\dagger c_1 - c_1^\dagger \alpha \rangle - 
\langle \beta^\dagger c_{\hat{L}} - c_{\hat{L}}^\dagger \beta \rangle \}
\:\:\:\: , 
\label{est.y1}
\eneq
\noindent
with $\langle \ldots \rangle$ denoting the average of the operator on the NESS. 
To implement the perturbative expansion in $g$, we now   introduce  
the eigenmodes of $ \hat{H}_B$ with energy $\lambda  \epsilon_k$ ($\lambda = \pm$),
 $\Gamma_{\alpha , \lambda}$, which are given by  
 \begin{eqnarray}
\Gamma_{ k , \lambda }                &=& \sum_{ j = 1}^{\hat{L}} ( \psi_{ j , k , \lambda} )^* c_j \nonumber \\
\Gamma_{0 ,\lambda} &=& \sum_{ j = 1}^{\hat{L}} (\psi_{j ,0 , \lambda})^* c_j 
\;\;\;\; , 
\label{est.2p}
\end{eqnarray}
\noindent
with the wavefunctions $\psi_{ j , k , \pm } , \psi_{ j , 0 , \pm}$ respectively provided in Eqs.(\ref{appe.a.4}) and 
Eqs.(\ref{appe.a.x.1}). The in-gap eigenmodes $\Gamma_{0 , \lambda}$ are present, or not, depending on 
whether the chain is in the topological, or in the trivial, phase. Inverting  
Eqs.(\ref{est.2p}), we obtain

\beq 
c_j  =  \sum_k \: \sum_{\lambda } \psi_{ j , k , \lambda} \Gamma_{k , \lambda} 
\:\:\:\: . 
\label{est.x1}
\eneq
\noindent
Eq.(\ref{est.x1}) allows us to express the right-hand side of Eq.(\ref{est.y1}) in terms
of the matrix elements of the covariance matrix $\theta$  \cite{Guimaraes2016},  according to  

\begin{eqnarray}
\langle \alpha^\dagger c_1 \rangle &=& \sum_{k , \lambda}  \: \psi_{ 1, k , \lambda} \langle \alpha^\dagger 
\Gamma_{k, \lambda}  \rangle =  \sum_k \: \sum_{\lambda } \psi_{ 1, k , \lambda} \theta_{\alpha ; (k , \lambda)} \nonumber \\
\langle \beta^\dagger c_{\hat{L}} \rangle &=& \sum_k \: \sum_{\lambda } \psi_{ L, k , \lambda} \langle \beta^\dagger 
\Gamma_{k, \lambda}  \rangle =  \sum_{k , \lambda}  \: \psi_{ L, k , \lambda} \theta_{\beta ; (k , \lambda)} 
\:\:\:\: . 
\label{est.x2}
\end{eqnarray}
\noindent
The leading nonzero contribution to the right-hand side of Eqs.(\ref{est.x2})  arises to  first order in $g$ and is given by 

\begin{eqnarray}
\theta_{\alpha ; (k, \lambda )} &=& \frac{ i g \psi_{ 1, k , \lambda} (\bar{f}_\alpha - \bar{f}_{k , \lambda} ) 
 }{\gamma - i \lambda \epsilon_k} \nonumber \\
\theta_{\beta ; (k, \lambda )} &=&  \frac{ i g \psi_{L , k , \lambda} (\bar{f}_\beta - \bar{f}_{k , \lambda}) 
}{\gamma - i \lambda \epsilon_k} 
\;\;\;\; , 
\label{est.x3}
\end{eqnarray}
\noindent
with   $\bar{f}_\alpha$ and $\bar{f}_\beta$ denoting  the Fermi distribution functions describing  respectively
 the left-hand and the right-hand bath level occupancies within the NESS.  To recover the large bias
regime  in the zero-temperature limit  we have to assume a negative (positive)
chemical potential $\mu_\alpha$ ($\mu_\beta$) for the left-hand (the right-hand) bath, respectively. This eventually implies 
$\bar{f}_\alpha = 1 , \bar{f}_\beta = 0$. With $\bar{f}_{k , \lambda}$  we denote the distribution function of the modes of the chain within 
the NESS. Due to the symmetry of $H_{\rm SSH}$ under the replacement $j \leftrightarrow L+1-j$, 
we obtain  $[\psi_{1,k, \lambda} ]^2 = [ \psi_{\hat{L} , k , \lambda} ]^2$, $\forall k , \lambda$.  
In this case,  our analogs of Eqs.(53,54) of \cite{Guimaraes2016} are 
trivially solved by $\bar{f}_{k , \lambda}^{\rm SSH} = \frac{ 1}{2} \{ \bar{f}_\alpha + \bar{f}_\beta \} = \frac{1}{2}$, independent of 
$k$ and $\lambda$. 
Finally, the particle-hole symmetry of 
the SSH chain at half-filling implies $ [\psi_{ 1 , k , + }]^2 = [\psi_{1 , k , -}]^2$ and $\epsilon_{k , + }= - \epsilon_{ k , - }$. Collecting the above results all together, and noting that when the full system is in the topological phase the subsystem $ \hat{H}_B$ 
is in the trivial one, the steady-state current in the NESS in the topological phase is given by          

\beq
I_{\rm NESS  }  =  \: \sum_{0 < k \leq \frac{\pi}{2}} \frac{\gamma  g^2 J_o^2  [ \psi_{1 , k , -} ]^2 }{ \gamma^2 + \epsilon_k^2 } 
\:\:\:\:,
\label{est.4}
\eneq
\noindent
with the sum taken over the values of $k$ that satisfy Eq.(\ref{appe.a.3}) with $J_e \leftrightarrow J_o$ and no contribution to the current arising from the localized, subgap 
modes. To 
access the trivial phase, we simply use again Eq.(\ref{est.4}), with just $J_e$ and $J_o$ exchanged with each other,
thus getting 

\beq
I_{\rm NESS }  =  \: \sum_{0 < k \leq \frac{\pi}{2}} \frac{\gamma  g^2 J_e^2  [ \psi_{1 , k , -} ]^2 }{ \gamma^2 + \epsilon_k^2 } 
\:\:\:\:,
\label{est.4bis}
\eneq
\noindent
with the sum over $k$ now computed over the solutions of the secular equation corresponding to propagating states 
(real $k$), as well as over the localized mode solution, corresponding to $k = \frac{\pi}{2} - i q$, with $q$ real and positive. 

\begin{figure}
 \center
\includegraphics*[width=1. \linewidth]{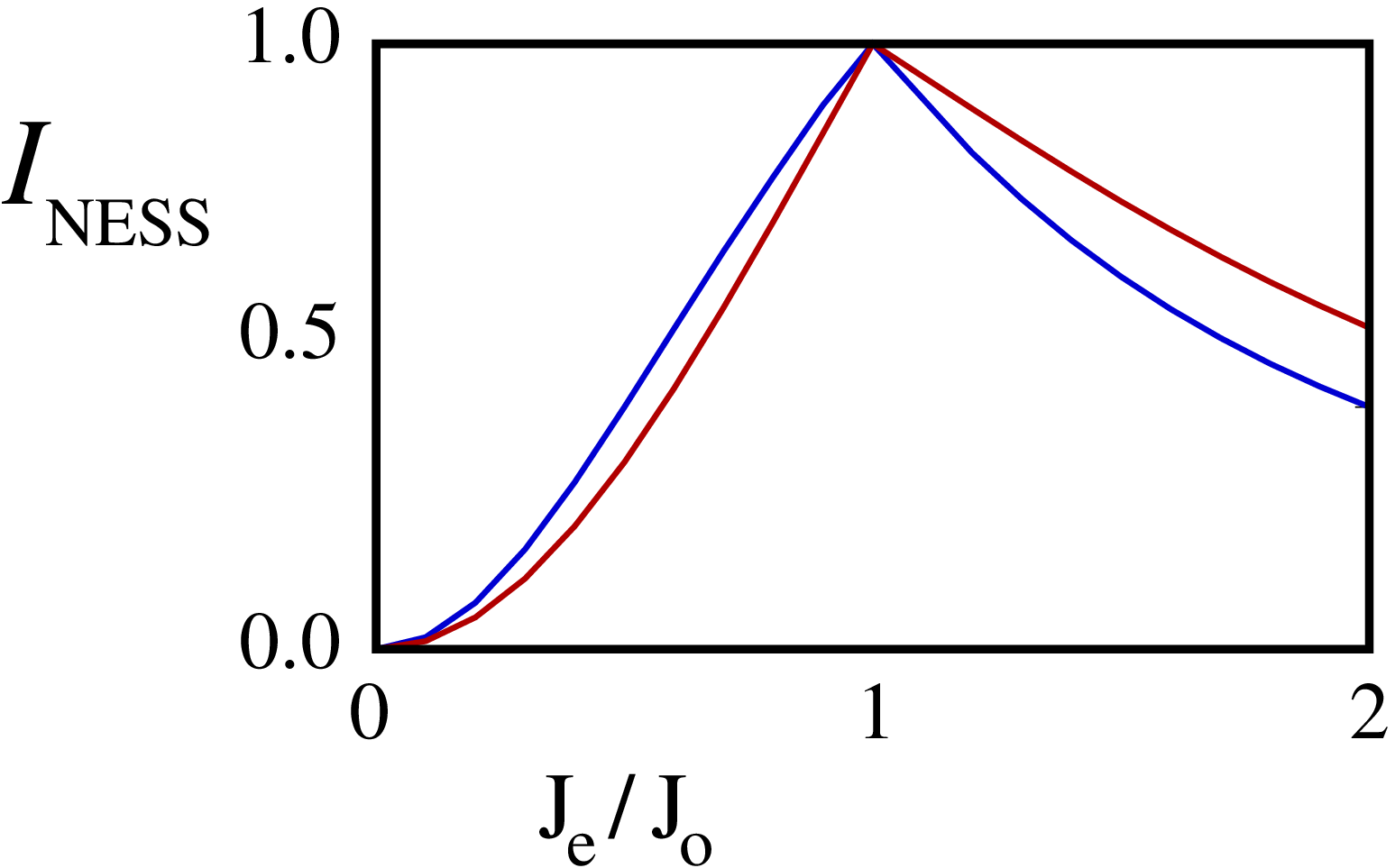}
\caption{$I_{\rm NESS}$ in an SSH chain  with $L=20$ as a function of $J_e/J_o$ at fixed $J_o=1$ and $\gamma /J_o = 1.5$, 
computed using the analytical 
expression in Eqs.(\ref{est.4},\ref{est.4bis}) (red curve) and by direct implementation of the Lindblad equation approach 
(blue curve). In both cases $I_{\rm NESS}$ has been normalized so that it is equal to 1 for 
$J_e/J_o=1$ (see main text for the corresponding discussion).
 } 
\label{current_comparation_topo}
\end{figure}

In Fig.\ref{current_comparation_topo} we plot $I_{\rm NESS}$ computed in an SSH chain with $L=20$ sites
with  the analytical formulas in 
Eqs.(\ref{est.4},\ref{est.4bis}) (red curve), as well as numerically calculated directly within the LE approach 
(blue curve),  by keeping $J_o$ fixed at 1 (reference energy value) and  $\gamma / J_o = 1.5$ and by varying $J_e$, for $0 \leq J_e / J_o \leq 2$. 
In general, $I_{\rm NESS}$ shows a maximum $I_M$ , as a function of $J_e/J_o$, at the topological phase transition, 
corresponding to the closure of the bulk gap of the chain at $J_e/J_o=1$. To get rid of the arbitrary parameter 
$g$ in Eqs.(\ref{est.4},\ref{est.4bis}), in drawing both curves we normalize $I_{\rm NESS}$ at the corresponding 
value of $I_M$. We note the  excellent agreement with each other, which evidences the due consistency 
 between the analytical and the fully numerical approach.          We also note that $I_{\rm NESS}$ keeps 
finite at any finite value of $J_e /J_o$.  

To rule out the possibility that (as it happens  when considering equilibrium dc transport  
throughout the SSH chain) this might be  some finite-size effect, below we provide a general argument 
proving that, at any finite values of $J_e$ and $J_o$ and in the absence of disorder, $I_{\rm NESS}$ keeps finite as 
$L \to \infty$.

\noindent
 To do so, we show that Eqs.(\ref{est.4},\ref{est.4bis}) are bounded from below and from above 
by  two quantities that remain finite in the thermodynamic limit. Indeed, we note that, as $L \to \infty$, 
 Eqs.(\ref{est.4},\ref{est.4bis})  yield 

\beq
I_{\rm NESS} \approx \frac{L}{4 \pi} \:\int_{-\pi}^\pi \: d k  \: \left\{  \frac{ \gamma  g^2 {\cal J}^2  }{\gamma^2 + \epsilon_k^2}  
\frac{                 [ 1 - \cos (2 k L )]}{
L - \frac{ \sin (kL) \cos [ k ( L + 2 ) ]}{ \sin ( 2 k ) } }\right\} 
\:\:\:\: , 
\label{est.1}
\eneq
\noindent
with ${\cal J} = {\rm min} \{  J_o , J_e \}$.  From Eq.(\ref{est.1}) we readily find that 

\beq
\frac{3}{( 3 \pi + 1 ) } {\cal I}_L  \leq I_{\rm NESS} \leq  {\cal I}_L 
\;\;\;\; , 
\label{est.2}
\eneq
\noindent
with 

\begin{eqnarray}
{\cal I}_L                &=& \frac{\gamma g^2 {\cal J}^2 }{2 \pi} \:\int_{-\pi}^\pi \: d k  \: \left\{ \frac{                [ 1 - \cos ( 2 k L )] }{\gamma^2 + \epsilon_k^2}  \right\} 
\nonumber \\
&=& \gamma g^2 {\cal J}^2  \left\{  \frac{ 1 -z_*^L }{ 2 \sqrt{ ( J_e^2 + J_o^2 + \gamma^2 )^2 - 4 J_e^2 J_o^2}}\right\} 
\:\:\:\: , 
\label{est.3}
\end{eqnarray}
\noindent
with 

\beq
z_* = - \frac{J_e^2 + J_o^2 + \gamma^2}{ 2 J_e J_o} + \sqrt{ \frac{ ( J_e^2 + J_o^2 + \gamma^2 )^2 - 4 J_e^2 J_o^2}{ 4 J_e^2 J_o^2} }
\:\:\:\: . 
\label{est.4x}
\eneq
\noindent
Eventually, Eqs.(\ref{est.3},\ref{est.4x}) imply that, as $L\to \infty$, we get 

\beq
{\cal I}_{L \to \infty} =    \left\{  \frac{\gamma g^2 {\cal J}^2}{ 2 \sqrt{ ( J_e^2 + J_o^2 + \gamma^2 )^2 - 4 J_e^2 J_o^2}}\right\} 
\:\:\:\: , 
\label{est.5}
\eneq
\noindent
which is always finite at any finite values of $J_e$ and $J_o$.
Due to disorder-induced localization, the situation fully changes as disorder is turned on, 
though, as we discuss next, strongly depending on the nature of the  disorder.

\subsection{ $I_{\rm NESS}$ and disorder-induced localization  in the presence of bond disorder}
\label{ines_bond}

In a one-dimensional conductor any amount of disorder 
would lead to complete localization of single electron states and to the corresponding suppression of 
long-distance charge transport in the scaling limit \cite{Anderson1958,Abrahams1979}. At variance,  disorder
induced by correlated impurities with a nontrivial internal structure 
can give rise to a mobility edge  that marks a metal-to-insulator   transition, even
in one-dimensional disordered conductors \cite{Dunlap1990,Phillips1991,Wu1992}.  
This is due to the fact that,  differently from what 
 happens with a  structureless  impurity located at a lattice site, an impurity with an internal
structure, such as a correlated dimer, allows, under pertinent conditions on the values of the 
various system parameters, for a single-particle tunneling through, with zero reflection amplitude and
just a total phase shift in the wavefunction  \cite{Wu1992}. 

That being stated,  following standard arguments \cite{Lambert1982}, one concludes that a chain of length $L$ hosts a number of localized 
states $N_{\rm loc}   \sim \sqrt{L}$. These states support conduction, though, in the thermodynamic 
limit, they determine a zero-measure set. The disappearance of these states marks the metal-to-insulator
phase transition.  As we argue here and in the following section, the current $I_{\rm NESS}$ provides
an effective way to detect the mobility edge   in the disordered SSH chain  
  \cite{Phillips1991b}.

To check where, and under what conditions, the mobility edge should arise in our system, we follow  
the main argument of \cite{Wu1992},   for both  bond, and dimer correlated  disorder. 
Specifically, we derive  the explicit expression for the reflection 
coefficient in the presence of the impurity, $r_k$,  as a function of the momentum $k$ and 
verify if, and under what conditions, $r_k = 0$, which marks the onset of the mobility edge 
  \cite{Wu1992}. In appendix \ref{solclean}
we derive $r_k$ in the SSH chain in the presence of a bond- and of a 
dimer-impurity. In the case of a bond impurity,  we  find 
that $r_k$ is given by

\beq
r_k =  \frac{ e^{ 4 i k + i \varphi_k} W ( 2-W  ) }{ - 1 + e^{ 2 i ( k + \varphi_k ) } ( 1-W )^2 } 
\;\;\;\; . 
\label{mose.7bisx}
\eneq

Eq.(\ref{mose.7bisx}) implies that, {\it regardless of $k$}, we can recover the condition $r_k = 0$ 
only if either $W=0$ (which corresponds to  the ``trivial'' case of absence of disorder, or $W=2$. 
In the scaling limit we therefore expect that bond disorder induces full localization of the
electronic states, except along the lines in parameter space corresponding to $W = 0 $ 
and to $W=2$ (note that $r_k$ is symmetric under $W \leftrightarrow 2 W$, so, 
we do expect a corresponding symmetry for  the phase diagram).  

To evidence  how this reflects in the  behavior of  $I_{\rm NESS}$, 
we have used the  LE approach to compute it
 in the presence of bond disorder, as a function 
of $W/J_o$ and of the ratio $J_e/J_o$ (using $J_o$ as unit of energy and accordingly 
setting it to 1) for  increasing values of $L$.  To  account for the disorder, 
at each set of values of $J_e/J_o ,W $ and $L$ we ensemble averaged  the 
 results for $I_{\rm NESS}$ over   $N=50$ realizations  of the bond disorder, with probability of having a  single altered bond set at
 $1 - \sigma = 0.5$. To optimize our procedure  by letting  the system operate at the optimal working point by maximizing 
 $I_{\rm NESS}$,   we repeated 
 the procedure of  \cite{Nava2021} and  eventually chose   $\gamma=2$. 
 
\begin{figure}
\includegraphics[scale=0.48]{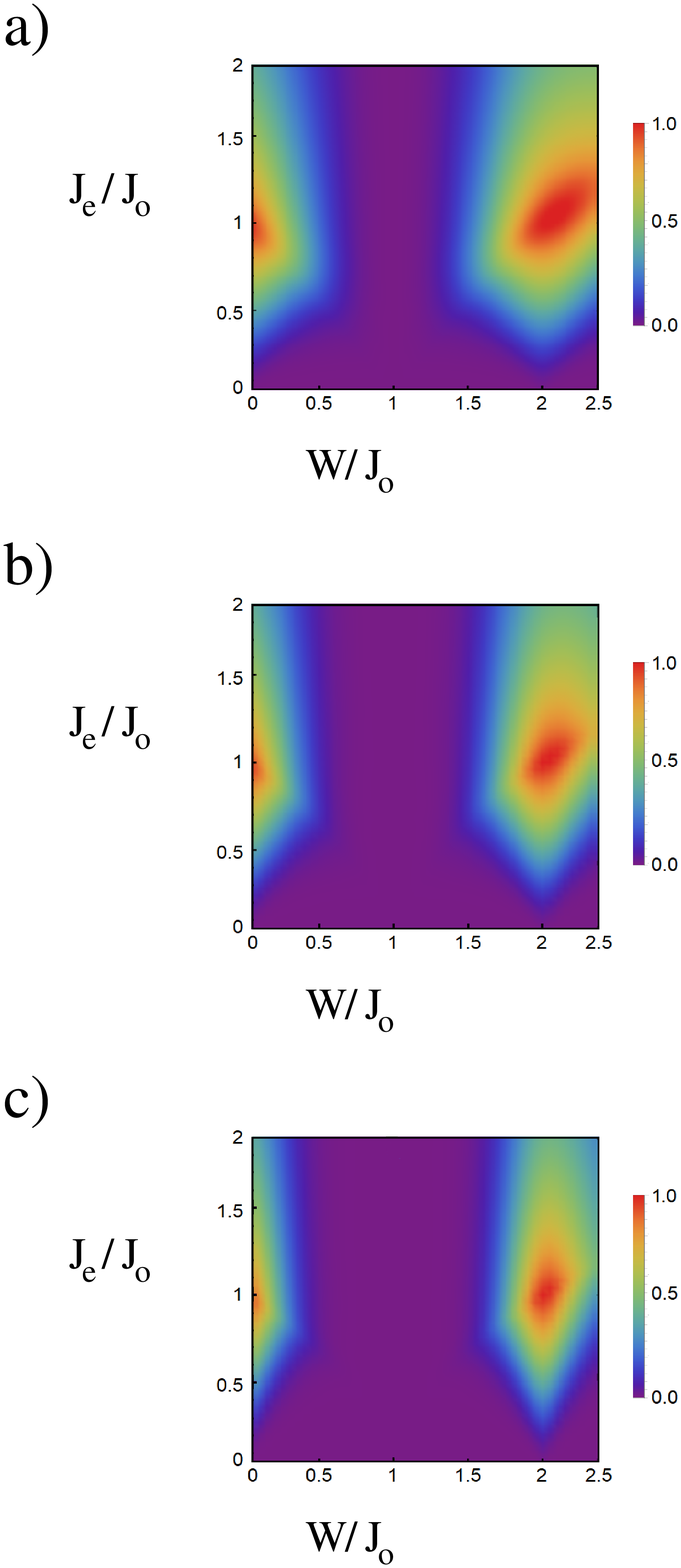}
\caption{\\
{\bf a):} $I_{\rm NESS}$ in an $L=20$ SSH chain with bond disorder in the hopping terms with $\sigma = 0.5$ (see Eq.(\ref{model.3})),
 as a function of $W/J_o$ and of $J_e/J_o$, for $0\leq W/J_o  \leq 2.5$ and 
for $0 \leq J_e/J_o \leq 2$ ($J_o$ has been used as a reference energy and it has therefore been set to 1). 
The figure has been generated by averaging over $N=50$ different realizations of the bond disorder. The 
couplings to the Lindblad baths have been set so that $\Gamma_1 = \gamma_L = \gamma = 2$. 
The color code highlights the value of $I_{\rm NESS}$ (normalized at its maximum) as evidenced in the
figure; \\
{\bf b):} Same as in panel {\bf a)} but with $L=40$;\\
{\bf c):} Same as in panel {\bf a)} but with $L=80$. }
\label{fig:hopping_phase}
\end{figure}

  In Fig.\ref{fig:hopping_phase} we draw our result for $I_{\rm NESS}$ as a function of $W/J_o$ and of $J_e/J_o$, for $0\leq W \leq 2.5$ and 
for $0 \leq J_e/J_o \leq 2$, in a chain at increasing values of $L=20,40,80$  connected to two Lindblad baths in the large-bias regime, 
 with  couplings to the bath $\Gamma_1 = \gamma_L \equiv \gamma$. From the color code for the value of $I_{\rm NESS}$ we see 
 that, on one hand, the current is maximal over regions centered on the lines  $W/J_o = 0 , 2$, which is consistent with 
 the criterion of \cite{Wu1992} and with the result of Eq.(\ref{mose.7bis}). Moving across panels {\bf a)}, {\bf b)} and {\bf c}, 
 we also see that, as expected, the larger is $L$, the 
 more the region in parameter space over which $I_{\rm NESS}$ is appreciably $\neq 0$ shrinks over the lines $W/J_o = 0 , 2$.
 Anywhere else the system is insulating as $L \to \infty$.  
 
\begin{figure}
\includegraphics[scale=0.5]{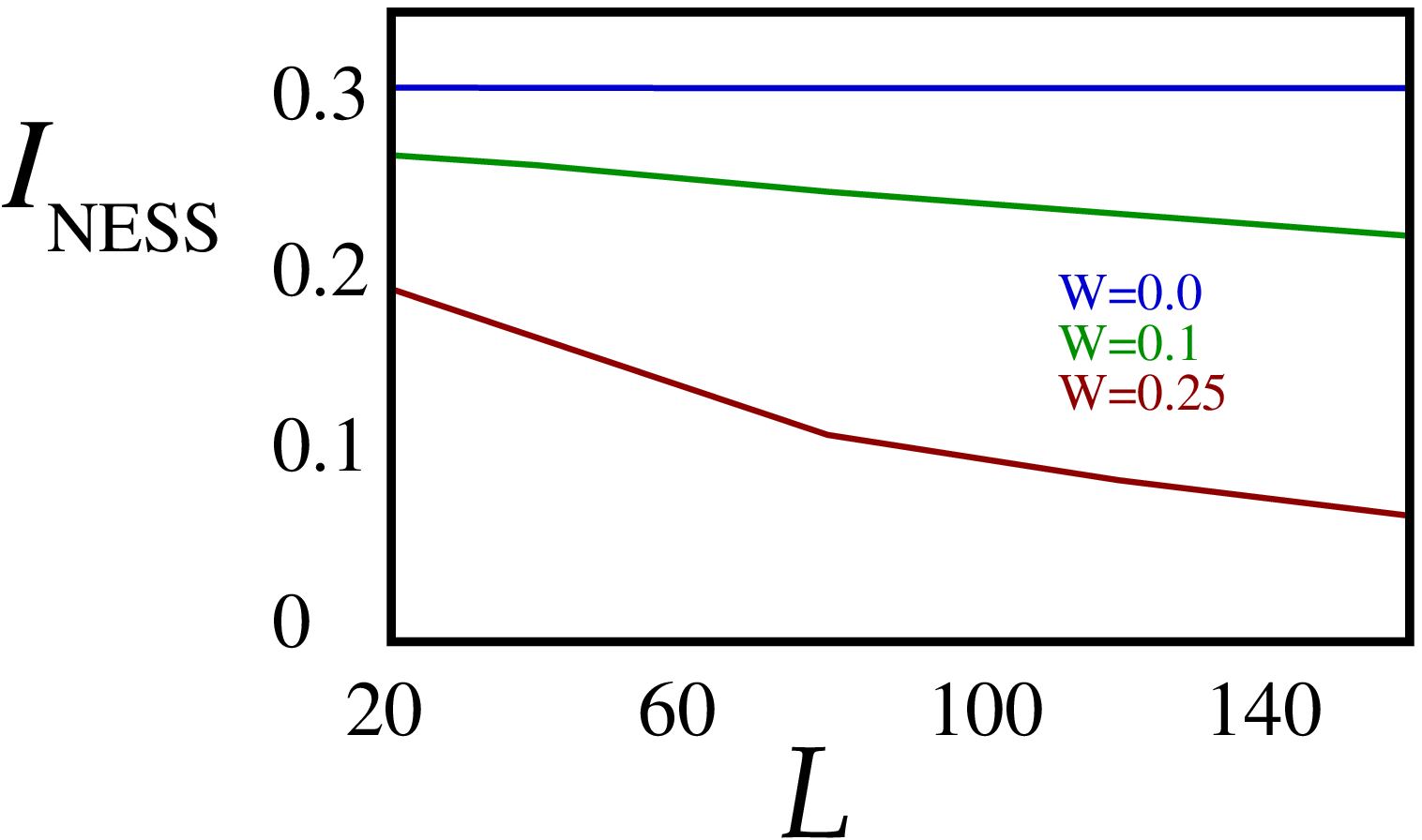}
\caption{$I_{\rm NESS}$ as a function of $L$ for $20 \leq L \leq 160$ computed in a SSH chain with $J_e/J_o=1.5$, 
with couplings to the Lindblad baths   set so that $\Gamma_1 = \gamma_L = \gamma = 2$, at 
a given realization of bond disorder, with $W=0.0$ (blue line), $W=0.1$ (green line), and $W=0.25$ (red line). }
\label{cross_check_bond}
\end{figure}

As an independent cross-check of our conclusions, in Fig.\ref{cross_check_bond} we plot $I_{\rm NESS}$ computed with 
the same parameters we used to draw Fig.\ref{fig:hopping_phase}, but at a single
realization of the disorder, as a function of $L$ for $20 \leq L \leq 160$, with $J_e/J_o=1.5$ and for $W/J_o = 0 , 0.1,0.25$. We clearly see that, 
while there is apparently no scaling at all (as it must be) for $W/J_o = 0$, in both the other cases 
$I_{\rm NESS}$ appears to scale toward 0 on increasing $L$.  We therefore conclude that, as $L \to \infty$, the  bond disorder suppresses 
$I_{\rm NESS}$ everywhere in parameter space, except along the vertical lines $W=0$ and $W =2$.

\noindent

\subsection{ $I_{\rm NESS}$ and disorder-induced localization  in the presence of dimer disorder}
\label{ines_dimer}

As we show in appendix \ref{solclean}, 
the reflection amplitude $r_k$  across a dimer impurity is given by

\beq
r_k = - \frac{ e^{ 4 i k + i \varphi_k} W (  W + 2 J_o  \cos ( k + \varphi_k) )}{ 
J_o^2 + 2 e^{ i ( k + \varphi_k ) } J_o W + e^{ 2 i ( k + \varphi_k ) } ( - J_o^2 + W^2 ) }  
\;\;\;\; . 
\label{mose.8x}
\eneq
\noindent
From Eq.(\ref{mose.8x}) we now see that $r_k=0$ if either $W=0$ (which, again, corresponds to the 
trivial limit of absence of disorder), or $W+ 2J_o \cos ( k + \varphi_k ) = 0$. This latter condition 
 corresponds to the result of \cite{Wu1992} for a dimer impurity,  once one respectively 
identifies $\epsilon_a$ and $V$ of that paper with our $W$ and $J_o$, and once one notes that, in 
our case, we have set $\epsilon_b$ of \cite{Wu1992} at 0 and have an additional contribution $\varphi_k$ in 
the argument of the cosine that disappears, once one sets $J_e = J_o$, as in \cite{Wu1992}.

The equation  $W+ 2J_o \cos ( k + \varphi_k ) = 0$ takes a solution $k_*$ for $0 \leq k \leq \pi$ only provided
that $ | W / 2 J_o | \leq 1$. If this is the case, over an $L$-site chain, $\sqrt{L}$ electronic states (out of the total of 
$L$ states) centered around $k_*$ are not localized by the disorder and, though, as $L \to \infty$, they contribute 
the total number of states by a zero measure set, they are nevertheless enough to support a conducting phase 
in the chain, thus determining the onset of a mobility edge  \cite{Dunlap1990,Wu1992}.        

To  evidence how this  affects   $I_{\rm NESS}$,  we  have generalized   Eqs.(\ref{est.4},\ref{est.4bis}) to the case in which a
single impurity is present in the chain. In this case, due to the lack of symmetry of the system under
the site-order inversion, $j \leftrightarrow L-j+1$,  Eqs.(\ref{est.4},\ref{est.4bis}) are substituted by 

\beq
I_{\rm NESS} = \sum_{0 \leq k \leq \frac{\pi}{2}} \left\{                \frac{ 2 \gamma g^2 {\cal J}^2 [                \psi_{ 1 , k , -}]^2 [\psi_{L , k , -}]^2 }{
( \gamma^2 + \epsilon_k^2 ) (  [                \psi_{ 1 , k , -}]^2 +  [\psi_{L, k , -}]^2 ) } \right\}
\:\:\:\: , 
\label{inb.2}
\eneq
\noindent
with ${\cal J} = {\rm min} \{ J_e , J_o \}$. At a value of $k$ corresponding to a localized state, at large enough system 
length $L$, either $\psi_{L,k,-}^2$ is exponentially suppressed compared to $\psi_{1,k,-}^2$, or vice versa, thus resulting into
an over-all suppression of the corresponding contribution to $I_{\rm NESS}$. At variance, within the window of delocalized 
states we obtain $ [         \psi_{ 1 , k , -}]^2 [\psi_{L , k , -}]^2 /  (  [                \psi_{ 1 , k , -}]^2 +  [\psi_{L, k , -}]^2 )  \approx (2 L)^{-1}$.
Neglecting also the dependence of the group velocity $v_k = \partial_k \epsilon_k$ on $k$ (which is appropriate far enough from 
the band edge), we eventually obtain 

\beq
I_{\rm NESS}          \approx                 \frac{ 2  g^2 {\cal J}^2 }{\pi} {\rm arctan} \left[          \frac{          \alpha \sqrt{L}}{\gamma}          \right]
\;\;\;\; , 
\label{inb.2x}
\eneq
\noindent
with $\alpha$ being a constant.  The right hand side of Eq.(\ref{inb.2x}) 
 takes a finite limit as $L \to \infty$. Clearly, an expression like Eq.(\ref{inb.2x}) is strictly related to the 
presence of a mobility edge within the allowed band of states, without which one would again obtain 
a complete, localization induced, exponential suppression of $I_{\rm NESS}$ as a function of $L$.

\begin{figure}
\includegraphics[scale=0.48]{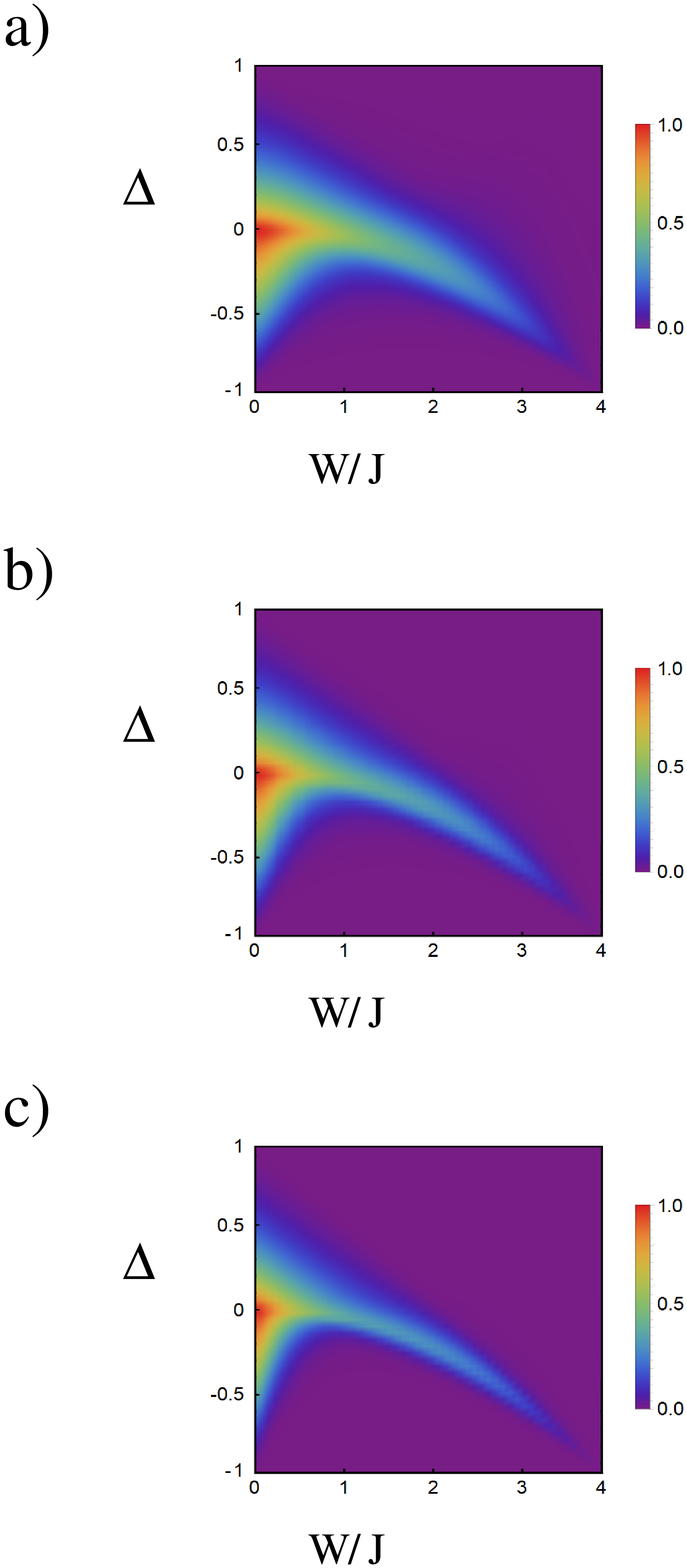}
\caption{\\
{\bf a):} $I_{\rm NESS}$ in an $L=20$ SSH chain with dimer disorder i  with $\sigma = 0.5$ (see Eq.(\ref{model.4})),
 as a function of $W/J $ and of $\Delta$, with $J =\frac{ J_e + J_o}{2}$ and $\Delta = \frac{ J_e - J_o}{J_e + J_o}$ and 
 for $0\leq W/J  \leq 4$ and 
for $-1 \leq \Delta \leq 1$.  The figure has been generated by averaging over $N=50$ different realizations of the dimer disorder. The 
couplings to the Lindblad baths have been set so that $\Gamma_1 = \gamma_L = \gamma = 2$. 
The color code highlights the value of $I_{\rm NESS}$ (normalized at its maximum) as evidenced in the
figure; \\
{\bf b):} Same as in panel {\bf a)} but with $L=40$;\\
{\bf c):} Same as in panel {\bf a)} but with $L=80$. }
\label{hopping_dimer}
\end{figure}

Along the derivation we performed in the case of bond disorder, we  again employed LE approach to compute $I_{\rm NESS}$ 
 in the   large-bias limit in the presence of dimer disorder of strength $W$ (see Eq.(\ref{model.4})), as a function 
of $W/J$ and of $\Delta$, for increasing values of $L$ and with $\Delta=\frac{\delta J}{J}$ where $J = \frac{J_e + J_o}{2}$ and $\delta J= \frac{J_e - J_o}{2}$. Again, to 
account for the effects of the disorder, at each value of $W/J$ and $\Delta$ we  ensemble averaged  
 $I_{\rm NESS}$ over $N=50$ realizations  of the dimer  disorder, with probability of having a  single altered dimer set at
 $1 - \sigma = 0.5$. As in drawing Fig.\ref{fig:hopping_phase}, we have chosen  $\Gamma_1 = \gamma_L \equiv \gamma = 2$.
 Aside from the trivial line $W=0$, we can clearly identify the light green/light blue region at finite $W/J$ with the 
 set of delocalized states around the mobility edge. Taking into account that, in drawing Fig.\ref{hopping_dimer} 
 we are choosing our energy units so that $J_e = 1 + \Delta$ and $J_o = 1- \Delta$, we  see that 
 the light-colored region is stuck at the line $W / J = 2 J_o / J = (2-2\Delta)/J$. As expected, on increasing $L$ 
 there is a mild sharpening of the light-colored regions, simply related to the corresponding reduction of the width
 of the region of extended states. Yet, the peak value of $I_{\rm NESS}$ is not suppressed as $L$ gets large, thus fully confirming 
 the persistence of the mobility edge in that  limit.

\begin{figure}
\includegraphics[scale=0.5]{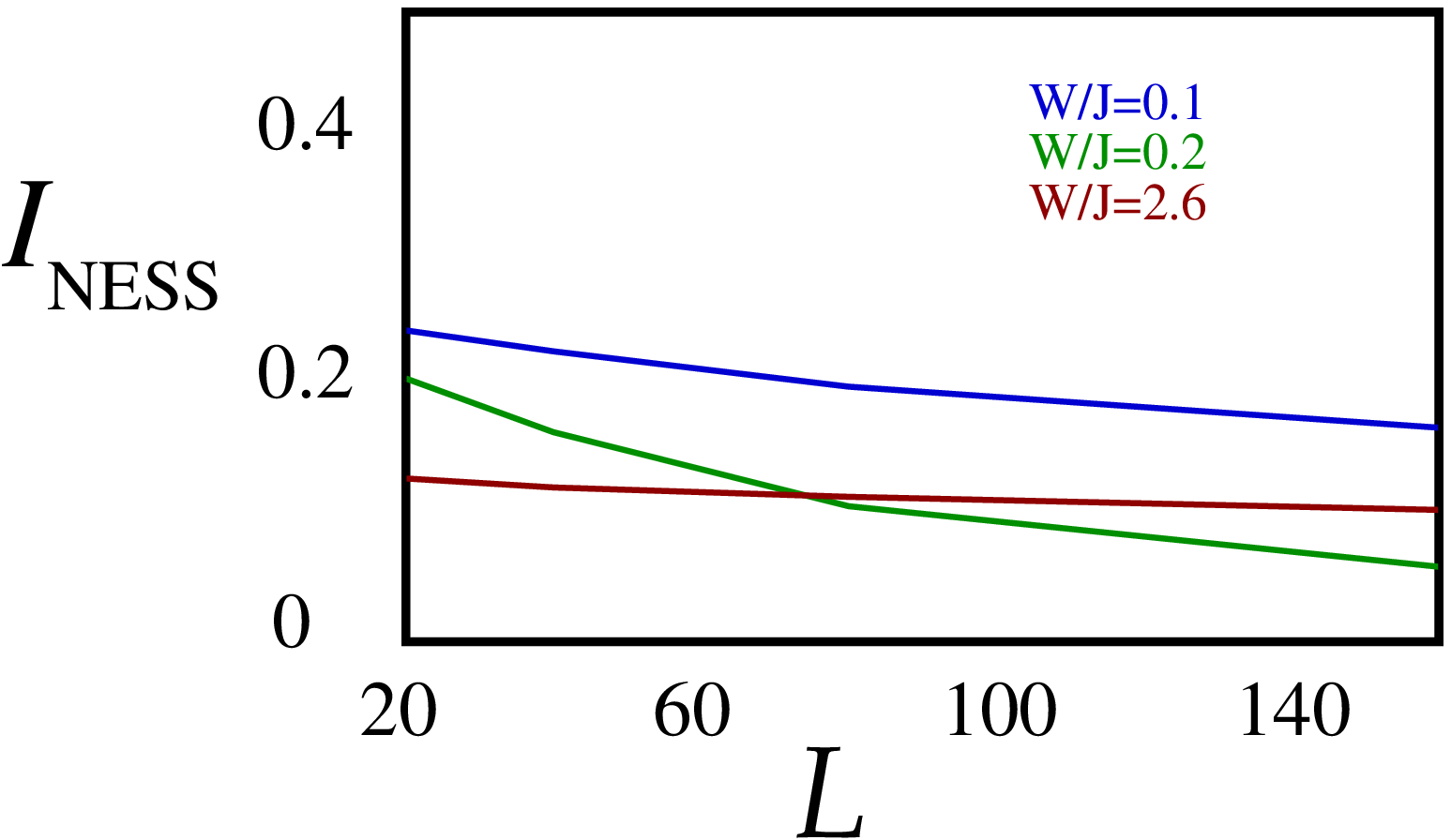}
\caption{$I_{\rm NESS}$ as a function of $L$ for $20 \leq L \leq 160$ computed in a SSH chain with $J_e=1+\Delta$, $J_o=1-\Delta$,
$\Delta=-0.4$, 
with couplings to the Lindblad baths   set so that $\Gamma_1 = \gamma_L = \gamma = 2$, at 
a given realization of dimer  disorder, with $W/J=0.1$ (blue line), $W/J=0.2$ (green line),
+99 and  $W/J=2.6$  (red line). }
\label{scaling_dimer}
\end{figure}

As a cross-check of our conclusions, in Fig.\ref{scaling_dimer} we show a sample of the scaling of $I_{\rm NESS}$ with $L$
at a specific point of the diagram in Fig.\ref{hopping_dimer}. Specifically, we draw $I_{\rm NESS}$ as a function of $L$
for $20 \leq L \leq 160$ for  $\Delta = -0.4$ and for $W/J=0.1$ (blue curve), $W/J=0.2$ (green curve) and $W/J=2.6$ (red curve), by taking the
system to the large bias limit, with the couplings between the chain and the bath chosen as in Fig.\ref{hopping_dimer}. 
At both  points at $W/J=0.1$ and $W/J=2.6$ (that is, close to the mobility edge)
we see practically no scaling of the current with the system size, which is consistent with the 
conclusion, evidenced by  Fig.\ref{hopping_dimer}, that at both points the chain is in the conducting phase. 
At variance, as soon as we move off the mobility edge, we see a suppression 
of $I_{\rm NESS}$ as $L$ increases, which is evident already   at $W/J=0.2$.

\noindent
To conclude, we have shown how, connecting the disordered SSH chain to two baths in the large bias limit and 
probing the steady-state current in the   NESS that sets in, provides an effective mean to map out the 
insulating and the conducting phases of the system, as well as the localization-delocalization phase transitions between
them. In particular, we have evidenced how the value of $I_{\rm NESS}$ is strongly sensitive 
to the evolution of the system through the mobility edge that arises in the presence of correlated dimer disorder. 
To complement the results of this section, in the following we introduce
the ``even-odd differential occupation'' as a mean to  map out the 
disorder-induced topological phase transitions in the system.

\section{The even-odd differential occupation}
\label{eod_0}

We now introduce the EOD
$\bar{\nu}$ as a collective property of the NESS of the chain and show   how it can be used  
 to detect  the topological/trivial nature of the NESS. 

When the system is at equilibrium, the onset of the topological phase  
corresponds to a nonzero value of the winding number $w$ \cite{Qi2008}.  Alternative
physical quantities sensible to the onset of nontrivial topology have been proposed, such as 
the charge polarization \cite{Qi2008,Kane2013}, and the dielectric polarization \cite{Aihara2020}. 
 Finally, the DAWN in Eq.(\ref{dawn.1}) has been proposed in the presence of 
disorder described by a potential that anticommutes with ${\bf \Gamma}$   \cite{Liu2022}.  

In the general case of a topological system in the presence of disorder, an effective 
mean to define the topological phase is by first ensemble-averaging over the disorder 
a quantity $\nu$ that is 0 in the trivial phase and $\neq 0$ (and quantized) in the topological phase
and then by defining a threshold value of $\nu$ above which the system is assumed to be in the topological 
phase \cite{Nava2017}.

To introduce the EOD and to illustrate its meaning, let us focus, for the time being, on  an $L$-site (with 
$L$ even) SSH chain  at equilibrium, in 
the   ``extreme'' limits. These are defined by respectively sending 
$J_e \to 0$ while keeping $J_o$ finite (the ``trivial'' extreme state), and by sending $J_o \to 0$, by keeping 
$J_e$ finite (the ``topological'' extreme state).   Denoting with 
  $H_{\rm Ext , Tr}$ and with  $H_{\rm Ext , To}$ the Hamiltonian of the system respectively describing 
  the trivial and the topological extreme state,    we 
 obtain 
 
 \begin{eqnarray}
 H_{\rm Ext , Tr} &=& - J_o \: \sum_{ r = 1}^{\frac{L}{2}} \{    c_{2r-1}^\dagger c_{2r} + c_{2r}^\dagger c_{2r-1} \} \nonumber \\
 H_{\rm Ext , To} &=& - J_e \: \sum_{ r = 1}^{\frac{L}{2}    - 1 } \{    c_{2r}^\dagger c_{2r+1} + c_{2r+1}^\dagger c_{2r }    \}    
 \:\:\:\: . 
 \label{trivial}
 \end{eqnarray}
 \noindent
 To diagonalize the Hamiltonians  in Eqs.(\ref{trivial}) we   rewrite them as

 \begin{eqnarray}
  H_{\rm Ext , Tr} &=& J_o \:\sum_{ r = 1}^\frac{L}{2}    \: \{    d_{ r , u}^\dagger d_{r , u } - d_{r , g}^\dagger d_{r , g }    \}
  \nonumber \\
 H_{\rm Ext , To} &=& J_e \: \sum_{ r = 1}^{\frac{L}{2} - 1 }    \: \{ f_{r , u}^\dagger f_{r , u} - 
 f_{r , g}^\dagger f_{r ,g } \}    
 \:\:\:\: ,
 \label{trivia.x1}
 \end{eqnarray}
 \noindent
 with, respectively 
 
 \begin{eqnarray}
  d_{r ,  g  }    &=& \frac{ c_{2r-1} + c_{2r}}{\sqrt{2}} \;\; , \; d_{r , u }    =  \frac{ c_{2r-1} - c_{2r}}{\sqrt{2}}  \nonumber \\ 
  f_{r , g  } &=& \frac{ c_{2r} + c_{2r+1}}{\sqrt{2}}  \;\; , \;   f_{r , u }  =  \frac{ c_{2r} - c_{2r+1}}{\sqrt{2}} 
       \:\:\:\: . 
 \label{trivia.x2}
 \end{eqnarray}
 \noindent
  Defining the empty state  $ | {\bf 0}    \rangle$ so that 
$ c_j | {\bf 0}    \rangle = 0$,   $ \forall j = 1 , \ldots , L$,   
 from Eqs.(\ref{trivia.x1},\ref{trivia.x2}) we readily see that,  
     in the trivial  extreme limit, the   groundstate $ | {\rm Tr} \rangle$ is uniquely 
 given by 
 
 \beq
 | {\rm Tr} \rangle = \prod_{    r = 1}^\frac{L}{2}    \: d_{r, g}^\dagger | {\bf 0} \rangle 
 \;\;\;\; .
 \label{trivia.7}
 \eneq
 \noindent
At variance, in the    topological extreme limit, the groundstate is twofold degenerate:   in this case, 
two orthogonal  groundstates  $ | L , {\rm To}    \rangle$ and $ | R , {\rm To}    \rangle$ are given by 
 
 \begin{eqnarray}
 | L , {\rm To}    \rangle &=& c_1^\dagger \prod_{ r = 1}^{\frac{L}{2}    - 1 }    f_{r , g }^\dagger | {\bf 0}\rangle \nonumber \\
 | R , {\rm To}    \rangle &=& c_L^\dagger \prod_{r = 1}^{\frac{L}{2}    -1 }    f_{ r ,g }^\dagger | {\bf 0}    \rangle 
 \:\:\:\: . 
 \label{trivia.8}
 \end{eqnarray}
 \noindent
 The two states in Eq.(\ref{trivia.8}) correspond to   the zero-energy Dirac fermion localized at 
either endpoint of the chain  \cite{Semenoff2006,Meier2016}.  It is worth stressing that this is drastically 
 different from what happens with the superconducting Kitaev chain,  where a zero-energy Dirac fermion
  can only be built as a superposition of the two Majorana modes at the endpoints of 
the chain (and is, therefore, nonlocal in real space).

 To discriminate between the states in Eqs.(\ref{trivia.7},\ref{trivia.8}) we define the 
  EOD $\bar{\nu}$  as the average of the operator ${\bf \Gamma}$ in Eq.(\ref{chiral.def}),  
 that is, $\bar{\nu} =  {\rm Tr} [ {\bf \Gamma} \rho ]$. Basically, 
$\bar{\nu}$ measures
the net average occupancy of the odd sites minus the one of  the even sites 
of the chain. 

 At equilibrium, when $T=0$, $\bar{\nu}$ reduces to the groundstate 
average of ${\bf \Gamma}$. In the extreme limits discussed above, 
 $ |{\rm Tr} \rangle , | L , {\rm  To} \rangle$, and $ | R , {\rm To} \rangle$ are
all eigenstates of ${\bf \Gamma}$ with eigenvalues respectively equal to 0, 1, and -1.  
In general, when computing a thermodynamical average using
 the density matrix, we expect that, being degenerate in energy,   $| L , {\rm To}    \rangle$ and 
 $| R , {\rm To}    \rangle$ equally weight for the final result, thus eventually 
 implying  $\bar{\nu}  = 0$.  Therefore, in  order to have a nonzero $\bar{\nu}$,   
 we have to  ``favor'' one of the two states with respect 
 to the other. As we are going to argue in the following, connecting the chain to two Lindblad baths in 
 the large-bias limit does perfectly accomplish this task. 
 
\begin{figure}
\includegraphics[scale=0.46]{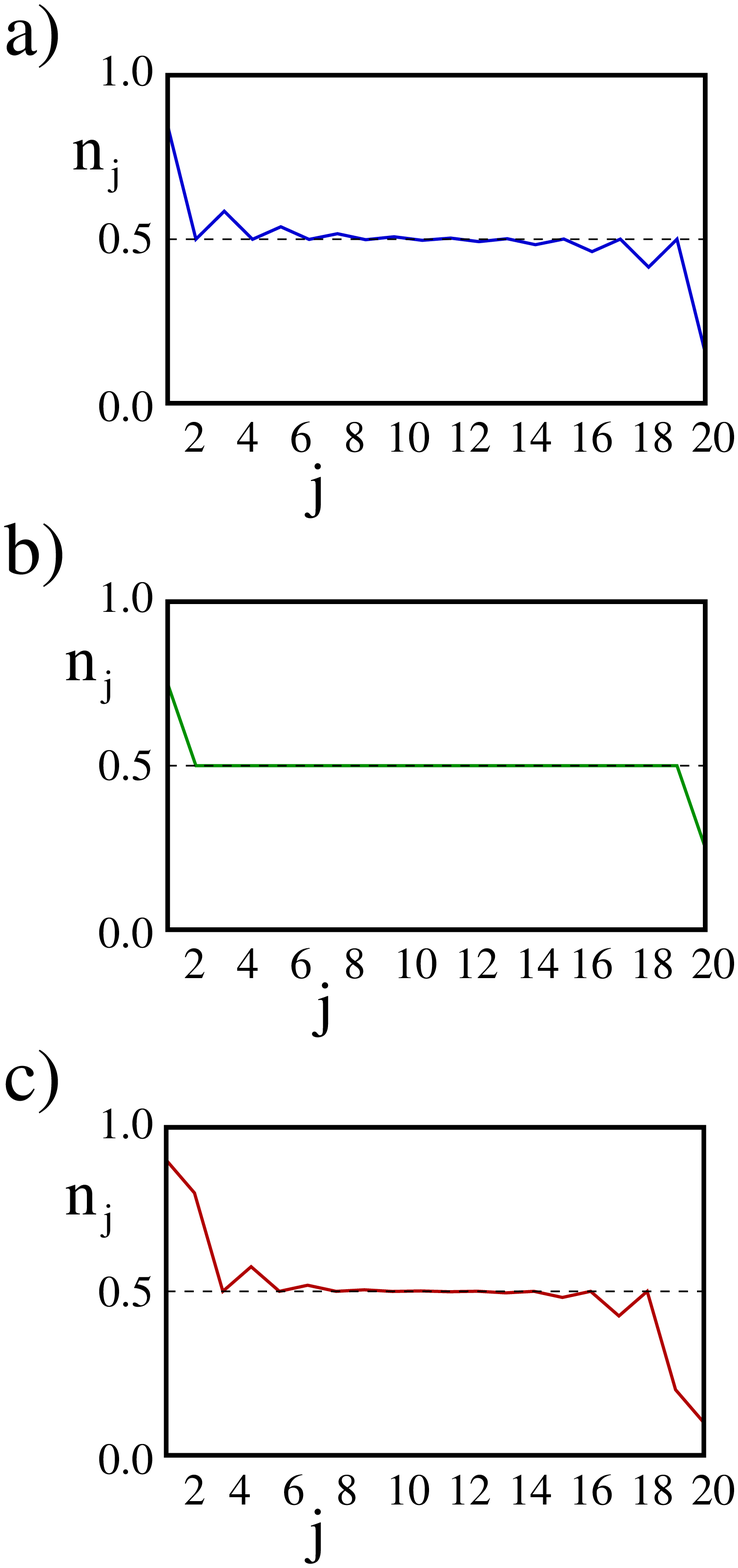}
\caption{{\bf a)} $n_j$ as a function of $j$ in an $L=20$ SSH chain with $J_e=1.5$ and $J_o=1$, connected to two 
baths in the large bias limit, with $\Gamma_1 = \gamma_L = 2$; \\
{\bf b)} $n_j$ as a function of $j$ in an $L=20$ SSH chain with $J_e=1$ and $J_o=1$, connected to two 
baths in the large bias limit, with $\Gamma_1 = \gamma_L = 2$; \\
{\bf c)} $n_j$ as a function of $j$ in an $L=20$ SSH chain with $J_e=0.5$ and $J_o=1$, connected to two 
baths in the large bias limit, with $\Gamma_1 = \gamma_L = 2$. }
\label{density_real}
\end{figure}

  To motivate our choice, in Fig.\ref{density_real},   we draw the real space 
 density $n_j$ over an $L=20$ SSH chain connected to Lindblad baths, computed, respectively, 
for $J_e=1.5,J_o=1$ (panel {\bf a)}, corresponding to  the topological phase), for $J_e = J_o = 1$ (panel 
{\bf b)},  corresponding to the topological phase transition), and for $J_e=0.5,J_o=1$ (panel {\bf c)}, corresponding 
to the topologically trivial phase). In all three cases we have considered the chain without disorder and 
in the large bias limit,  with $\Gamma_1 = \gamma_L  = 2$ (in units of $J_o$). Fig.\ref{density_real}{\bf  b)} 
is a particular case of the density distribution in a one-dimensional conducting chain at half-filling connected to two Lindblad baths 
in the large bias limit \cite{Nava2021}. The density is uniform and constantly equal to 1/2 everywhere, except at the two end
sites of the chain, where, due to the coupling to the baths, the density is shifted upwards or downwards from the otherwise uniform
value. $n_j$ as a function of $j$ shows a similar trend in Figs.\ref{density_real}{\bf a)} and {\bf c)}, except that now 
there is a staggering that modulates the decay from the boundary values to the uniform ``bulk'' value 1/2. In the topological 
case (panel {\bf a)} the first and the second oscillation, starting from either endpoint and moving toward the bulk of the chain, 
are large and in opposite directions: this makes $n_1 - n_2$ and $n_{L-1} - n_L$ provide by large the leading contribution to 
$\bar{\nu}$. In addition, $n_2-n_3$ ($n_{L-2}-n_{L-1}$) have opposite sign with respect to $n_1-n_2$ ($n_{L-1}-n_L$):
all this is expected to make the EOD ``large'' and pretty close to either $+1$, or $-1$, depending on the sign of the bias between the baths. 
At variance,  in the trivial  case (panel {\bf c)}  $n_1 - n_2$ and $n_{2} - n_3$, as well as $n_{L-2} - n_{L-1}$ and 
$n_{L-1}-n_L$, are apparently smaller than in the topological case and, more importantly, they have the same sign, being 
comparable in size, as well. This is now expected to make the EOD small, and close to 0.

\begin{figure}
\includegraphics[scale=0.46]{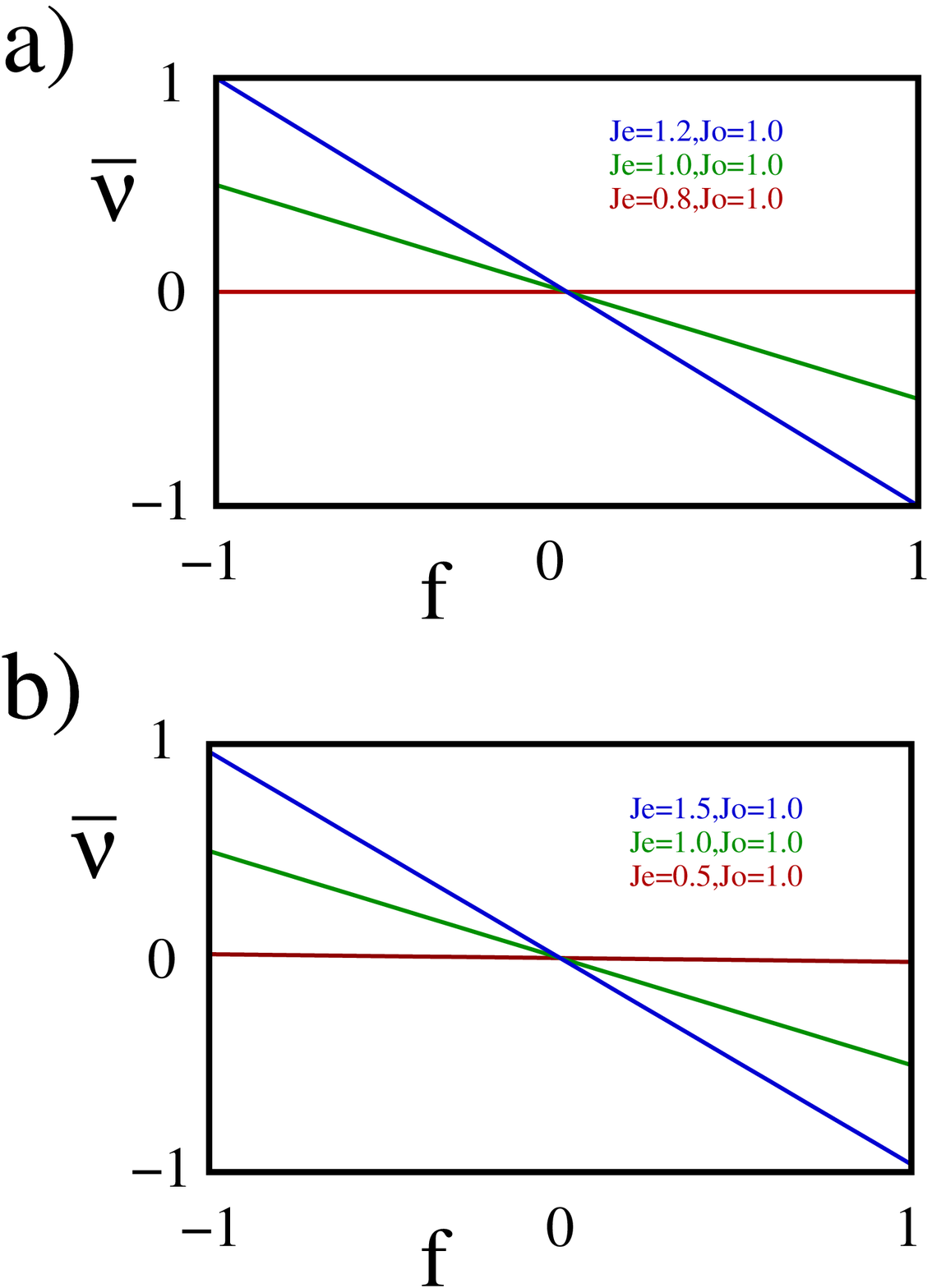}
\caption{{\bf a)} $\bar{\nu}$ as a function of $f$ in an SSH chain with $L=20$ sites, connected to two 
Lindblad baths characterized by $\Gamma_1 = \gamma_L = \frac{g}{2}                (1-f), \gamma_1 = \Gamma_L = \frac{g}{2}                ( 1 + f )$
(see Eq.(\ref{rates.1}) and the corresponding discussion for details) with $g=2$ and $- 1 \leq f \leq 1$, with 
$J_e=1.2,J_o=1.0$ (blue line), $J_e= J_o=1.0$ (green line),  $J_e=0.8J_o=1.0$ (red line); \\
{\bf b)} Same as in panel {\bf a)}, but with $J_e=1.5,J_o=1.0$ (blue line), $J_e= J_o=1.0$ (green line),  $J_e=0.5,J_o=1.0$ (red line). }
\label{sweeping_eod}
\end{figure}

To double-check our conclusion, we now   connect an SSH chain with $L=20$ sites  to two baths by its endpoints and set the parameters 
in Eq.(\ref{rates.1}) so that $\Gamma_1 = \gamma_L = \frac{g}{2}                (1-f), \gamma_1 = \Gamma_L = \frac{g}{2}                ( 1 + f )$, with 
$g=2$ and $- 1 \leq f \leq 1$ (so that the large-bias limits correspond to either $f=0$, or to $f=1$). 
In Fig.\ref{sweeping_eod} we plot  a sample of our results: in particular, in both panels of the figure we draw the 
``critical'' sweep, done by computing $\bar{\nu}$ at $J_e = J_o = 1.0$ and by varying $f$, as well as a sweep 
realized in the topological region ($J_e=1.2,J_o=1.0$ in Fig.\ref{sweeping_eod}{\bf a)},  $J_e=1.5,J_o=1.0$ in Fig.\ref{sweeping_eod}{\bf b)}),
and another one realized in the trivial  region ($J_e=0.8, J_o=1.0$ in Fig.\ref{sweeping_eod}{\bf a)},  $J_e=0.5,J_o=1.0$ in Fig.\ref{sweeping_eod}{\bf b)}).
Apparently, moving from the trivial to the topological region determines an increase in $\bar{\nu}$ by at least two order of 
magnitudes. In addition, in the large-bias limit and in the topological region $\bar{\nu}$ converges to the ``universal'' value $\pm 1$, depending 
on the sign of the bias.

\begin{figure}
\includegraphics[scale=0.46]{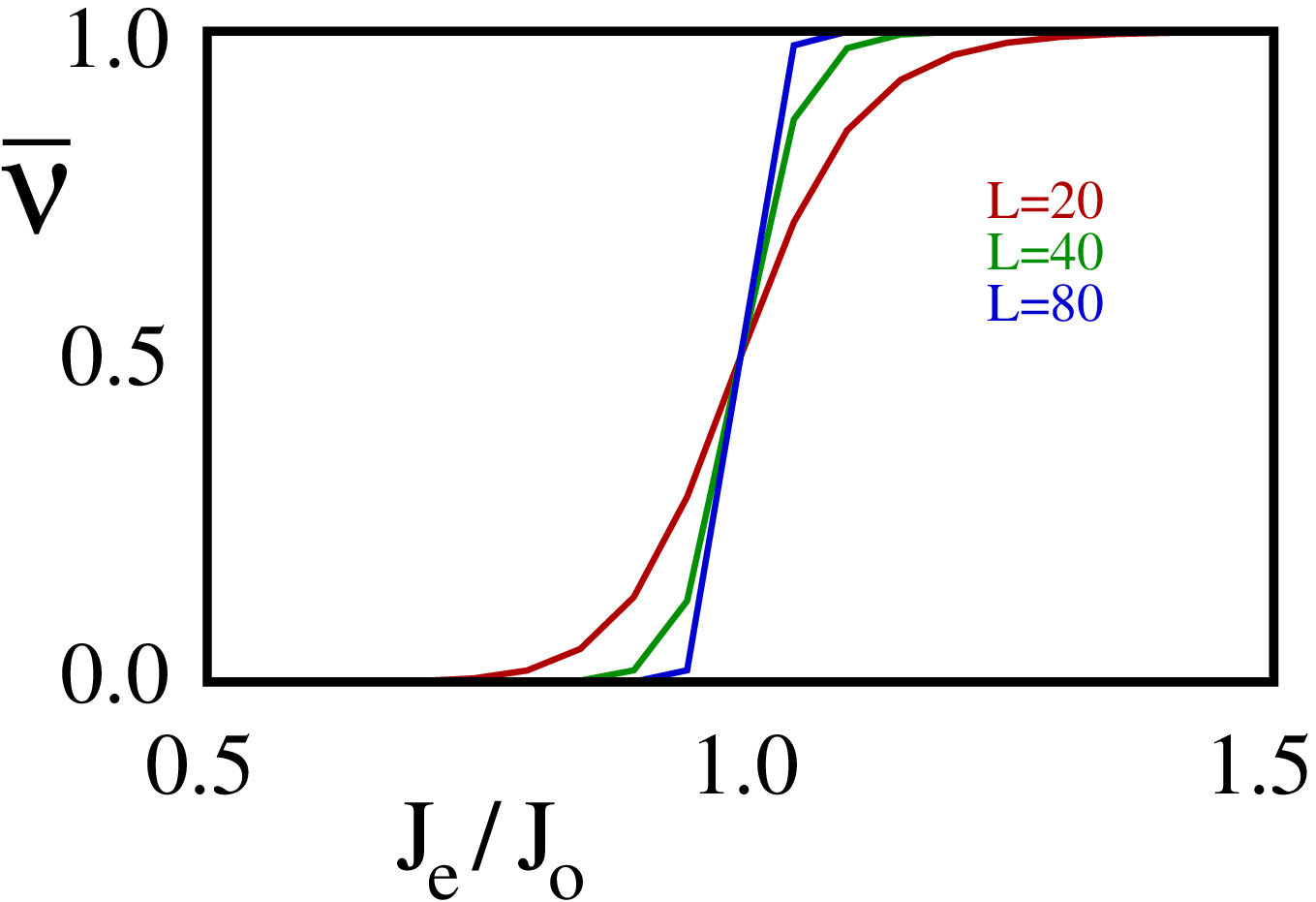}
\caption{$\bar{\nu}$ as a function of $f$ in an SSH chain  connected to two 
Lindblad baths in the large bias limit  ($\Gamma_1 = \gamma_L =g=2$, $\gamma_1 = \Gamma_L = 0$) 
as a function of $J_e/J_o$ for $0.5 \leq J_e/J_o \leq 1.5$, with the length of the chain 
 $L=20$ (red curve), $L=40$ (green curve), and $L=80$ (blue curve).  }
\label{eod_L}
\end{figure}

As a complementary analysis, we have also computed $\bar{\nu}$ in the large bias limit across the topological phase transition, for 
different values of the system length $L$. In Fig.\ref{eod_L} we plot our results for $\bar{\nu}$ as a function of $J_e/J_o$, with 
$J_o$ held fixed at 1, $\Gamma_1 =  \gamma_L = g = 2$,  $\gamma_1 = \Gamma_L = 0$, 
 and $-0.5 \leq J_e \leq 1.5$, for $L=20$ (red curve), $L=40$ (green curve), and $L=80$ (blue curve). We 
evidence the apparent switch of $\bar{\nu}$ from the value $\sim 0$ that it takes within the trivial region to the value $\sim 1$
that it takes in the large bias limit within the topological region. In addition, we appreciate how, the larger is $L$, the sharper 
is the crossover region between the two limiting values, thus suggesting that, for large enough $L$, $\bar{\nu}$ just switches 
between 0 and 1 (or vice versa) at the topological phase transition.

\noindent
Of course all our results of this section apply to the clean limit: we now discuss how they 
are affected by the presence of  disorder.

\subsection{The  even-odd differential occupation at nonzero bond disorder}
\label{eod_bond}

By construction, any realization of the bond disorder potential anticommutes with ${\bf \Gamma}$.
 This implies that all the nondegenerate 
states of the disordered Hamiltonian at fixed disorder   are grouped in pairs with opposite energies, 
with states in a pair connected to each other by   ${\bf \Gamma}$.  Thus, we expect that   
when pushing the chain to the large bias limit and probing   $\bar{\nu}$ at a given realization of 
the disorder,  either 
the system is in the topologically trivial phase and $\bar{\nu}  = 0$, or 
it is in the  nontrivial phase and, in  the large-bias limit, 
we obtain  $\bar{\nu} = \pm 1$ (depending on the sign of the applied bias): the 
relative  probability of the two results  is a function of only the disorder strength.  
Counting how many times, over several realizations of the disorder,  $|\bar{\nu} | = 1$, 
 evidences whether the disorder itself is strong enough to destroy the topological phase, 
or not.   

\begin{figure}
\includegraphics[scale=0.4]{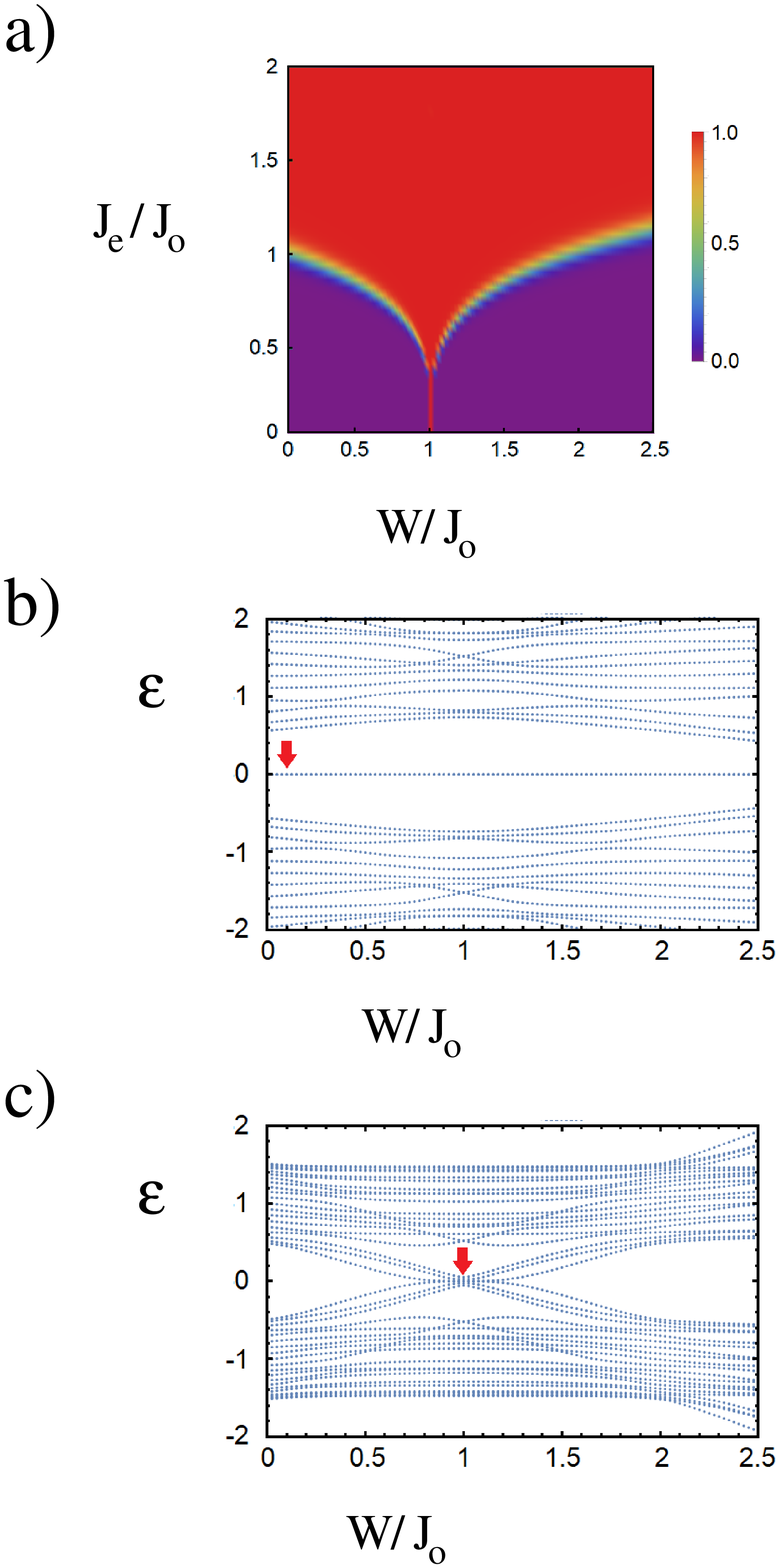}
\caption{\\ {\bf a):} $\bar{\nu}$ computed in an SSH chain with $L=40$ in the large bias limit with 
$\Gamma_1 = \gamma_L = 2$ and $\gamma_1 = \Gamma_L = 0$,
at a single configuration of bond disorder as a function of $W/J_o$ and of $J_e/J_o$, with 
$J_o$ (in the clean limit) used as the reference energy scale for the system. Red regions correspond to $\bar{\nu} = \pm 1$, 
deep purple regions to $\bar{\nu} = 0$;\\
{\bf b):} Energy levels of the chain as a function of $W/J_o$ at $J_e/J_o=1.5$ (the ingap states are evidenced by 
the red arrow);\\
{\bf c):} Same as in panel {\bf b)} but with $J_e/J_o = 0.5$.      }
\label{eod_bond_1}
\end{figure}

Here we first verify that, at a single realization of the disorder with variable disorder strength $W$, 
$\bar{\nu}$ is either $\pm 1$, or 0, depending on whether the model with the effective $J_o$ renormalized by 
the disorder lies within the topological, or the trivial, phase. We consider  an $L=40$ site chain connected to two 
external baths in the large bias limit, with   $\Gamma_1 = \gamma_L = 2$  and  $\gamma_1 = \Gamma_L = 0$.
The disorder is described by the distribution in Eq.(\ref{model.3}), with varying $W$.  In Fig.\ref{eod_bond_1}{\bf a)}           we show a color plot
of the measured EOD as a function of both   $W/J_o$ and  $J_e/J_o$, with $J_o$, normalized 
to 1, used as the reference energy scale.  The red area corresponds to $\bar{\nu} = 1$,    the deep purple area to $\bar{\nu} = 0$.
In Fig.\ref{eod_bond_1}{\bf b)}  and {\bf c)} we show a scatter plot of the 
energy levels of the chain as a function of $W/J_o$            at $J_e/J_o$ fixed and equal to 1.5 and to 0.5, so that, in the 
clean limit, the chain lies within respectively the topological, and the trivial phase. We evidence with a red arrow the 
emerging ingap states in both cases, which are the fingerprint of a topologically nontrivial system. We clearly see
that the red (deep purple) region of Fig.\ref{eod_bond_1}{\bf a)} corresponds to regions with a pair of ingap states 
(no ingap states) in the level diagram of the system, thus supporting  the effectiveness of using the EOD as a tool to monitor the
onset of the topological phase.

\begin{figure}
\includegraphics[scale=0.46]{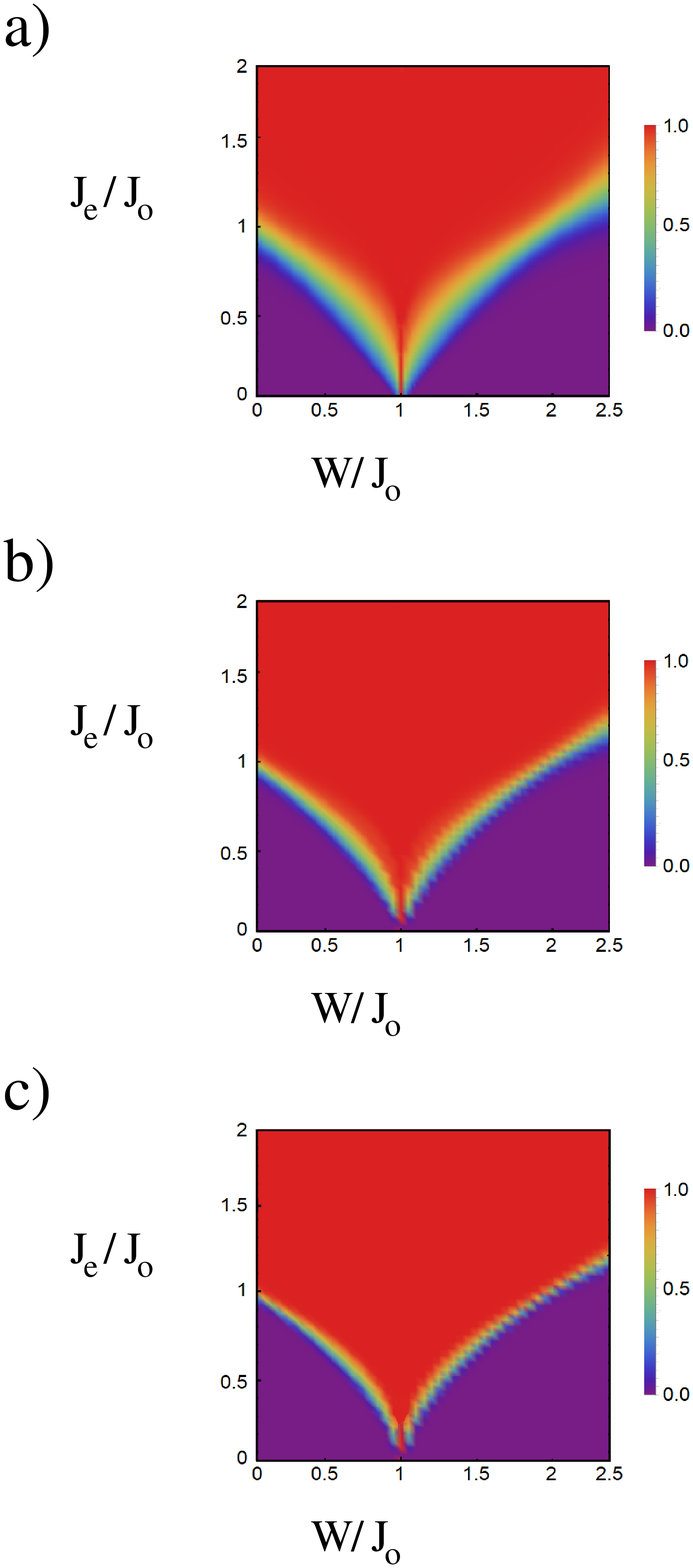}
\caption{\\
{\bf a):}$\bar{\nu}$ computed in an SSH chain with $L=20$ in the large bias limit $\Gamma_1 = \gamma_L = 2$ 
and  $\gamma_1 = \Gamma_L = 0$, and  
with chiral disorder by averaging over 50 realizations of the disorder by varying $W/J_o$ and $J_e/J_o$; \\ 
{\bf b):} Same as in panel {\bf a)}           but with $L=40$;\\
{\bf c):}          Same as in panel {\bf a)} but with $L=80$.    }
\label{comparison_grif_bond}
\end{figure}

 To check the  sharpening of the transition on increasing 
the chain length $L$, we have computed $\bar{\nu}$ 
 in a disordered chain in the large bias limit with 
$\Gamma_1 = \gamma_L = 2$ and $\gamma_1 = \Gamma_L = 0$, by ensemble averaging over $N=50$
realizations of the bond disorder extracted using ${\cal P}_{\rm b} [ J_o ]$ in Eq.(\ref{model.3}) at increasing 
$L$, for $L=20$, $L=40$, and $L=80$. 
We present the corresponding result in Fig.\ref{comparison_grif_bond} by using the same style as 
in Fig.\ref{eod_bond_1}{\bf a)} but, of course, by reporting now the ensemble-averaged results. The sharpening 
of the (light colored) transition region is apparent, which enforces our interpretation of the nature of the phase 
transition.

Having shown the effectiveness of the EOD in monitoring the onset of the topological phase in the disordered system 
taken at large bias with bond disorder, we now repeat the same analysis in the case  in which the disorder 
is realized with a random distribution of dimer impurities. 

\subsection{The even-odd differential occupation at nonzero dimer disorder}
\label{eod_bond}

Differently from alternative means, such as the DAWN, which does not work if 
the disorder does not anticommute with ${\bf \Gamma}$  \cite{Mondragon2014,Liu2022},
 as we show next, the EOD  also works with the  (``nonchiral'') dimer disorder. 
 
\begin{figure}
\includegraphics[scale=0.4]{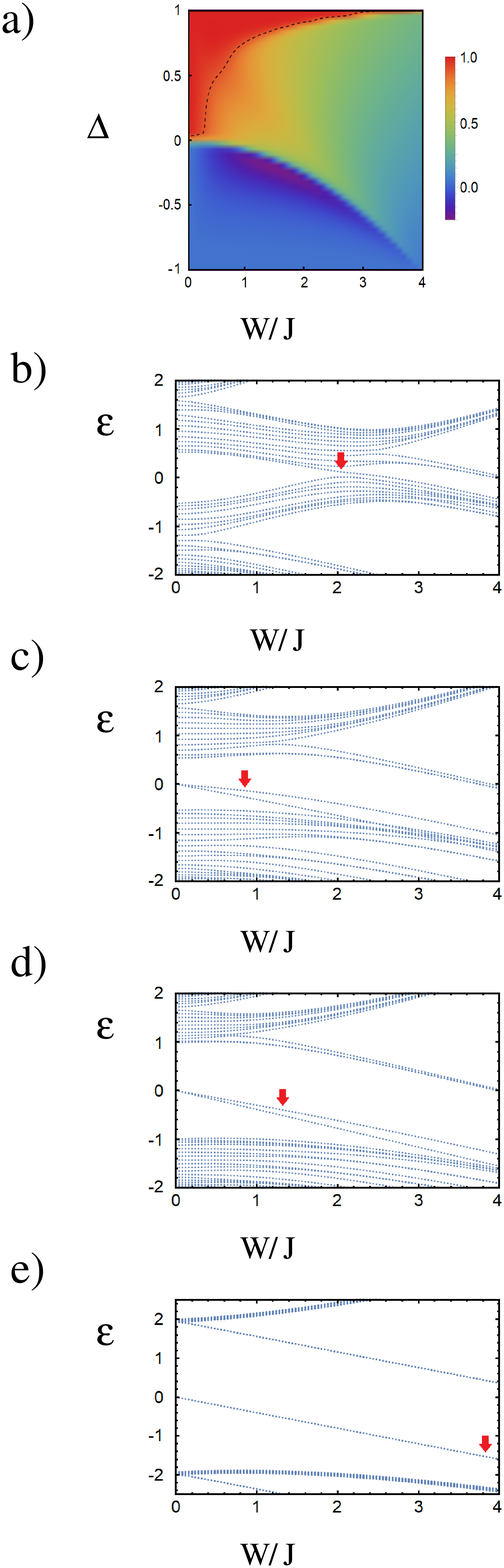}
\caption{\\ {\bf a):}$\bar{\nu}$ computed in an SSH chain with $L=40$ in the large bias limit with $\Gamma_1 = \gamma_L = 2$ 
and $\gamma_1 = \Gamma_L = 0$  
at a single configuration of dimer disorder as a function of $W / J $ and of $\Delta=\frac{J_e-J_o}{J_e+J_o}$. Red regions correspond to $\bar{\nu} = \pm 1$. The dashed black 
line marks the points at which $\bar{\nu} = \bar{\nu}_* = 0.90$;\\
{\bf b):} Energy levels of the chain as a function of $W/J$ at $\Delta=-0.25$;\\
{\bf c):} Same as in panel {\bf b)} but with $\Delta=0.25$ (now the ingap states are evidenced by a 
red arrow); \\
{\bf d):} Same as in panel {\bf c)} but with $\Delta=0.50$; \\
 {\bf e):} Same as in panel {\bf c)} but with $\Delta=0.95$. }
\label{eod_dimer_1}
\end{figure}

First of all let us note that, since the disorder potential does no longer anticommute with ${\bf \Gamma}$,
 we do not longer expect (in the large-$L$ limit) a sharp borderline between the region with $\bar{\nu} = \pm 1$ and 
the region with $\bar{\nu}          = 0$. This is apparent in   Fig.\ref{eod_dimer_1} {\bf a)}, where 
  we plot $\bar{\nu}$ computed in an SSH chain with $L=40$ at a single 
realization of the dimer disorder, as a function of the disorder strength normalized to the 
average hopping $W / J$, with $J=\frac{ J_e+J_o}{2}$,  and of $\Delta=\frac{\delta J}{J}$ with 
$\delta J = \frac{ ( J_e - J_o )}{2}$.  From  Fig.\ref{eod_dimer_1} {\bf a)} we see that, on increasing $W$ at fixed $\Delta (>0)$ 
(that is, starting from the topological phase, in the clean limit), the EOD smoothly 
evolves from the red region at small disorder (with $\bar{\nu} = \pm 1$) to a yellow, then green, then light blue region, 
corresponding to progressively (and continuously) decreasing values of $| \bar{\nu}|$.  To relate this behavior  to 
a possible suppression of the topological phase,    we again consider the energy levels at the same realization of the disorder as a function of 
$W /J$. An important point, here, is that, since, differently from the bond disorder, introducing dimer impurities does 
alter the over-all chemical potential, in order to have a common energy reference at any value of $W/J$ we systematically
subtract from the computed energy eigenvalues the extra over-all chemical potential determined by the 
added impurity (impurities). Doing so, we obtain the plots in Fig.\ref{eod_dimer_1}{\bf b)},{\bf c)},{\bf d)},{\bf e)}, where we 
draw the energy levels as a function of $W/J$ at a single realization of the disorder and for, respectively, 
$\Delta J = -0.25,0.25,0.50,0.95$. In panels {\bf c)},{\bf d)},{\bf e)},  we highlight
 with a red arrow the pair of ingap states, that are the 
fingerprint of the topological phase, to  which the system goes back  as $W \to 0$. Of course, we see 
no ingap states in panel {\bf b)}, as, in this case, the system goes to the trivial phase as $W \to 0$.
 It is worth stressing that, due to the peculiar nature of dimer disorder, increasing 
$W$ moves the ingap states toward either the upper, or the lower, gap edge (depending on 
the sign of $W$). Over a finite-$L$ disordered chain it is therefore natural to assume that the
ingap states  ``merge'' with the other states when their energy, measured with respect to the closer gap
edge, equates the average spacing between the other states. 
Using this criterion, we see that,  on increasing $W$,  the ingap doublet   merges  with the 
other states at a value of the disorder strength which, by looking at the color plot in panel {\bf a)}, we 
realize to correspond to a color crossover from red to yellow-green. On the numerical side, this means 
that $\bar{\nu}$ become $<1$ at the point in which the ingap states merge with the other states and 
the topological phase is lost. Accordingly, we can use $\bar{\nu}$ to define a border for the topological 
phase as the line at which the EOD takes a conventional value, which we choose to be $\bar{\nu}_* = 0.90$
(other choices such as, for instance, $\bar{\nu}_*=0.85$ do not substantially affect the shape of the 
borderline). In Fig.\ref{eod_dimer_1} {\bf  a)} we draw as a dashed black line 
the borderline of the topological phase that we define in this way.

\begin{figure}
\includegraphics[scale=0.5]{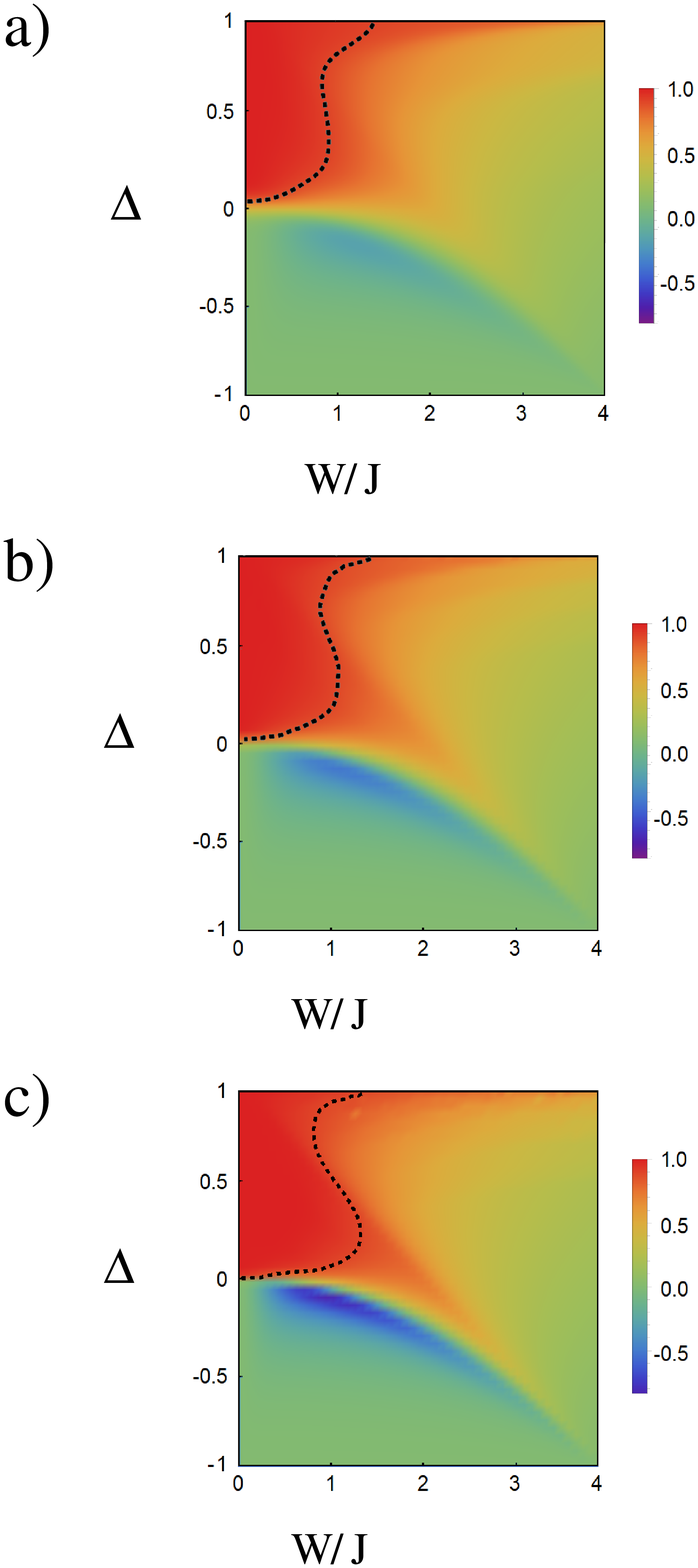}
\caption{\\
{\bf a):}$\bar{\nu}$ computed in an SSH chain with $L=20$ in the large bias limit with $\Gamma_1 = \gamma_L = 2$ 
and $\gamma_1 = \Gamma_L = 0$, in the presence of dimer disorder by   averaging over 50 realizations of the disorder by 
varying $W / J$ and 
$\Delta$. As in Fig.\ref{eod_dimer_1}, the dashed black 
line marks the points at which $\bar{\nu} =  0.90$; \\ 
{\bf b):} Same as in panel {\bf a)}           but with $L=40$; \\
 {\bf c):} Same as in panel {\bf a)}           but with $L=80$.   }
\label{comparison_grif_dimer}
\end{figure}

\noindent
To infer if, and to what extent, increasing the system size $L$ affects the global behavior of 
the EOD as a function of $W/J$ and of $\Delta$ (and, therefore, the shape and the position of 
the borderline of the topological phase),  we have  computed $\bar{\nu}$ 
by ensemble averaging over $N=50$ realizations of the dimer disorder in an SSH chain taken to the large bias limit
with $\Gamma_1 = \gamma_L = 2$  and $\gamma_1 = \Gamma_L = 0$, by varying both $W/J$ and $\Delta$ and for 
$L=20,40$ and $80$. We draw the corresponding diagrams in    Fig.\ref{comparison_grif_dimer}: by comparing the 
plots in Fig.\ref{comparison_grif_dimer} {\bf a)}, {\bf b)}, and {\bf c)} (corresponding to $L=20,40,80$)
 with each other, we
see no appreciable difference in the behavior of  $\bar{\nu}$ in the topological region. 
Thus, on one hand we conclude that the dimer disorder tends to suppress the topological 
phase of the SSH chain, except for a region originating from the line $W=0,0<\Delta$ and 
extending toward the right. On the other hand, the border of such a region, which we conventionally
attributed to $\nu = \nu_* = 0.90$, is not substantially affected by increasing $L$: 
 contrary to what happens with bond disorder, with dimer disorder 
there is no sharpening of the transition on increasing  $L$.

\noindent
To conclude, we evidence how, despite the apparent difficulty of even defining the borderline of the topological region in 
the presence of dimer disorder, taking the (open) chain in the large bias limit and going through a synoptic comparison of 
the variation of the EOD $\bar{\nu}$ as a function of the system parameters and of the evolution of the system levels as 
a function of the disorder strength at fixed $\Delta$, allows us to  mark  the border of, the topological region even in this case.
 
Combining the results of the EOD and of $I_{\rm NESS}$ for both kinds of disorder we have discussed above,  allows for
mapping   out the complete phase diagram of the SSH chain connected to two Lindblad reservoirs in the large bias limit, as a function 
of the chain parameters as well as of the strength of the disorder. 

\section{Conclusions}
\label{conclusions}

We have applied the LE method to derive the phase diagram of an open SSH chain connected to 
two external baths in the large bias limit, in the presence of bond  and of dimer disorder.  Biasing the external baths
has allowed us to stabilize a NESS, characterized by a steady current $I_{\rm NESS}$. Whether 
$I_{\rm NESS}$ flows to 0, or to a finite value, in the limit of large chain length $L$, tells us whether the system 
is fully localized by disorder, or not.
 
Our approach just needs a simple transport measurement, combined 
with an appropriate scaling analysis, to map out the localization/delocalization transition in the disordered chain.
In particular, our method has proven to be  effective in evidencing the expected onset of the mobility edge which, 
in the case of dimer disorder, marks the opening of the window of delocalized states in the otherwise fully localized 
band  \cite{Dunlap1990,Phillips1991,Wu1992}. 

We have used the (ensemble averaged) EOD occupancy
to distinguish between  topologically trivial and nontrivial phases. From the presence of ingap states,
in the spectrum of the SSH chain at a given realization of disorder, 
we have been able to relate  a nontrivial topological phase to an average EOD  $\bar{\nu} = \pm 1$ (depending on the sign of the applied bias).
 Computing the EOD poses no constraints on 
the symmetries of the disorder potential: as we show, we can compute the EOD regardless of whether the 
disorder anticommutes with ${\bf \Gamma}$, or not. Thus,   using the EOD, we can circumvent the   limitations of 
 alternative quantity, such as the DAWN  \cite{Mondragon2014,Liu2022}. 
 
 Of course,  a meaningful definition and 
implementation of the EOD requires that the system is taken to the out-of-equilibrium 
regime, into an appropriate NESS. We therefore conclude that, by driving the disordered SSH chain toward the 
large bias regime and by performing the appropriate measurements on the NESS that  sets in, it 
is possible to map out the complete phase diagram of the system, both for what concerns the localization/delocalization 
phase transition, as well as the topological phase transition, in the presence of disorder.

While, to define and illustrate the application of our method, in this paper we limited ourselves to a 
well-known and widely discussed physical system, such as the disordered SSH chain, 
there are no particular limitations in extending our technique to more complex and/or less investigated 
models. In this direction, an intriguing perspective we intend to pursue is applying our method to 
a disordered  SSH chain with long-range single-electron hopping and/or interaction. In this case,  a derivation of 
$I_{\rm NESS}$ as a function of the system parameters should allow for probing the   emergence of 
 mobility edge(s) in the system, as a consequence of the long-range nature of the correlations \cite{Rozhin2022}. 
Moreover, monitoring $\bar{\nu}$ should provide us with detailed informations about the onset of 
unconventional topological phases, which are in general expected to emerge 
 in systems with long-range correlations \cite{Vodola2014,Giuliano2018}. 
 
 Other possible applications 
 should concern, for instance, topological Kondo systems at junctions of quantum wires
  \cite{Beri2012,Tsvelik2013,Giuliano2013,Giuliano2020,Giuliano2020_2,Buccheri2022,Giuliano2022,Guerci2021}. 
 Finally, as an additional  possible development of our work we 
 believe it is worth mentioning the use of a multi-site Lindblad bath, so to attempt to smoothly interpolate between the
 linear response, ``Landauer-B\"uttiker like'' regime, and the large bias limit, so to connect the behavior of 
 $I_{\rm NESS}$ to the current response of the system in the linear regime. 
 
 \vspace{0.4cm}
 
{\bf Acknowledgements:} A. N. and D. G  acknowledge   financial support  from Italy's MIUR  PRIN project  TOP-SPIN 
(Grant No. PRIN 20177SL7HC).

\appendix 

\section{Derivation of Eq.(\ref{eq:HF-master})}
\label{derk}

  In this appendix we illustrate  the derivation of Eq.(\ref{eq:HF-master}) of the main text for 
 the matrix ${\cal C} ( t )$  by focusing, as a specific example, on the equation for the matrix element
 ${\cal C}_{1,1} ( t )$.   
 
 The starting point are the canonical anticommutation relations between the lattice single-fermion operators
 
\begin{eqnarray}
&& \{ c_{i}^{\dagger},c_{j} \}    =\delta_{i,j} \nonumber \\
&&  \{ c_{i}^{\dagger},c_{j}^{\dagger} \}    = \{ c_{i},c_{j} \} =0
 \: . \label{appe.X.1} 
\end{eqnarray}
\noindent
By writing the system Hamiltonian $H$ as $H = \sum_{i,j=  1}^L c_i^\dagger{\cal H}_{i,j}  c_j$ and by employing 
Eqs.(\ref{appe.X.1}), we obtain  (term-by-term):
 
 \begin{enumerate}
 
\item \begin{eqnarray}
&& \langle [H,c_{1}^{\dagger}c_{1} ] \rangle    =   \langle  [\sum_{i,j = 1}^L 
c_{i}^{\dagger}{\cal H} _{i, j}c_{j},c_{1}^{\dagger}c_{1} ] \rangle \nonumber \\
 && = \sum_{j\neq1} {\cal H}_{1,j}  \langle  [c_{1}^{\dagger} c_{j},c_{1}^{\dagger}c_{1} ] \rangle 
 +\sum_{i\neq1}{\cal H}_{i,1} \langle  [c_{i}^{\dagger}c_{1},c_{1}^{\dagger}c_{1} ] \rangle \nonumber \\
 && = \sum_{j\neq1} {\cal H} _{1,j} \langle -c_{1}^{\dagger}c_{j} \rangle +\sum_{i\neq1}{\cal H}_{i,1} \langle c_{i}^{\dagger}c_{1} \rangle \nonumber \\
  && = -\sum_{j}  [{\cal H}_{1,j} {\cal C}_{1,j}-{\cal H}_{j,1}{\cal C}_{j,1} ]
   = [{\cal H}^\top ,{\cal C} ]_{1,1} \; .
   \label{appe.X.2} 
\end{eqnarray}
\noindent 
 
 \item  \begin{eqnarray}
&& \Gamma_{1} \langle c_{1}c_{1}^{\dagger}c_{1}c_{1}^{\dagger} \rangle  = \nonumber \\
&& \Gamma_{1} \langle c_{1}\left(1-c_{1}c_{1}^{\dagger}\right)c_{1}^{\dagger} \rangle =
\Gamma_{1} \langle c_{1} c_{1}^{\dagger} \rangle   \nonumber \\ 
 && =\Gamma_{1}-\Gamma_{1} \langle c_{1}^{\dagger}c_{1} \rangle \; . 
 \label{appe.X.3}
\end{eqnarray}
\noindent

\item \begin{eqnarray}
&& -\frac{\Gamma_{1}}{2} \langle  \{ c_{1}c_{1}^{\dagger},c_{1}^{\dagger}c_{1} \}  \rangle \nonumber \\
 && =-\frac{\Gamma_{1}}{2} \langle c_{1}c_{1}^{\dagger}c_{1}^{\dagger}c_{1}+c_{1}^{\dagger}c_{1}c_{1}c_{1}^{\dagger} \rangle  
  =0 \; .
  \label{appe.X.4} 
\end{eqnarray}
\noindent

\item \begin{eqnarray}
&& \Gamma_{L}\ \langle c_{L}c_{1}^{\dagger}c_{1}c_{L}^{\dagger} \rangle -\frac{\Gamma_{L}}{2} 
\langle \left\{ c_{L}c_{L}^{\dagger},c_{1}^{\dagger}c_{1}\right\}  \rangle  \nonumber \\ 
 && =\Gamma_{L} \langle c_{1}^{\dagger}c_{1}c_{L}c_{L}^{\dagger} \rangle 
 -\frac{\Gamma_{L}}{2} \langle c_{1}^{\dagger}c_{1}c_{L}c_{L}^{\dagger}+c_{1}^{\dagger}c_{1}c_{L}c_{L}^{\dagger} \rangle  \nonumber \\
&&  =0 \; .
  \label{appe.X.5} 
\end{eqnarray}
\noindent

\item \begin{equation}
\gamma_{1} \langle c_{1}^{\dagger}c_{1}^{\dagger}c_{1}c_{1} \rangle =0 \;.
\label{appe.X.6} 
\end{equation}
\noindent

\item  \begin{eqnarray}
&& -\frac{\gamma_{1}}{2} \langle \{ c_{1}^{\dagger}c_{1},c_{1}^{\dagger}c_{1} \}  \rangle \nonumber \\ 
 && =-\frac{\gamma_{1}}{2} \langle c_{1}^{\dagger}c_{1}c_{1}^{\dagger}c_{1}+
 c_{1}^{\dagger}c_{1}c_{1}^{\dagger}c_{1} \rangle  =-\gamma_{1} \langle c_{1}^{\dagger}c_{1} \rangle  \; .
 \label{appe.X.7}
\end{eqnarray}
\noindent

\item  \begin{eqnarray}
&& \gamma_{L} \langle c_{L}^{\dagger}c_{1}^{\dagger}c_{1}c_{L} \rangle -\frac{\gamma_{L}}{2} 
 \langle  \{ c_{L}^{\dagger}c_{L},c_{1}^{\dagger}c_{1} \}  \rangle \nonumber \\
 && =\gamma_{L} \langle c_{L}^{\dagger}c_{L}c_{1}^{\dagger}c_{1} \rangle 
 -\frac{\gamma_{L}}{2} \langle c_{L}^{\dagger}c_{L}c_{1}^{\dagger}c_{1}
 +c_{L}^{\dagger}c_{L}c_{1}^{\dagger}c_{1} \rangle \nonumber  =0 \: . 
 \label{appe.X.8}
\end{eqnarray}
\noindent
\end{enumerate}
Adding the terms listed above all together  we obtain  Eq.(\ref{eq:HF-master})  of our manuscript for 
${\cal C}_{1,1} (t)$. A straightforward generalization of the above procedure yields the equations for 
a generic matrix element ${\cal C}_{i,j} ( t )$.  
 
\section{Solution of the SSH model in the clean limit}
\label{solclean}

In this appendix, we review the derivation of the single-particle wavefunctions for the SSH chain in the absence 
of disorder. In particular, we first solve the time-independent Schr\"odinger equation over an $L$-site open
chain and derive the corresponding wavefunctions, which we use in section \ref{transport_NESS} to analytically 
compute $I_{\rm NESS}$. Then,  we compute the scattering amplitudes across a single impurity, both
in the case of a bond impurity and of a dimer impurity. The corresponding result for the reflection coefficient 
is what we use in sections \ref{ines_bond} and \ref{ines_dimer}    to analyze  the impurity induced localization 
in the NESS.

\subsection{SSH chain with open boundary conditions}
\label{ssh_open}

We now review the derivation of the eigenvalues and of the eigenfunctions of an $L$-site SSH 
chain with open boundary conditions.  This is described by  the Hamiltonian $H_0$, given by 

\beq
H_0  =  -   \sum_{ j = 1}^{L-1} J_{j,j+1}  \{  c_j^\dagger c_{j+1} + c_{ j +1}^\dagger c_j \}      
\:\:\:\: , 
\label{appe.1.1}
\eneq
\noindent
with 
 
\beq
J_{j,j+1} = \Biggl\{ \begin{array}{l}     J_o \;\; (j \: {\rm odd} ) \\
J_e \;\; (j \;{\rm even}) 
\end{array} 
\;\;\;\; , 
\label{addi.appe.1}
\eneq
\noindent
supplemented with open boundary conditions at both the endpoints of the chain.
Taking into account the staggering due to the term $\propto \delta J$, we consider 
energy eigenmodes in the form 

\beq
\Gamma_\epsilon = \sum_{ j = 1}^L \{ u_j + (-1)^j v_j \} c_j 
\equiv \sum_{ j =     1}^L \psi_j c_j 
\;\;\;\; , 
\label{appe.1.2}
\eneq
\noindent
with  the open boundary conditions $u_{j=0} + v_{j=0} = 
u_{j = L+1}           - v_{ j = L+1}           = 0$ (note that, as in the main text here we are 
assuming that $L$ is even). On imposing 
$[  \Gamma_\epsilon , H_0 ] = \epsilon \Gamma_\epsilon$, we obtain the lattice,
time-independent Schr\"odinger equation for the wavefunctions $u_j , v_j$ in the form 

\begin{eqnarray}
\epsilon u_j &=& - \frac{J_e + J_o}{2}  \{  u_{ j +1} + u_{j-1}\}           
+    \frac{ J_e - J_o}{2}   \{   v_{j+1} - v_{j-1}    \}  \nonumber   \\
\epsilon v_j &=& -  \frac{ J_e - J_o}{2}  \{    u_{j+1} - u_{j-1} \}           
+ \frac{J_e + J_o}{2}          \{  v_{j+1}  + v_{j-1} \}   \:\:\:\: ,     \nonumber \\
 \label{appe.1.3} 
\end{eqnarray}
\noindent
with $1 <  j <  L$.
Assuming a solution of  the form 

\beq
\left[  \begin{array}{c}           u_j \\ v_j \end{array}           \right]           
= \left[ \begin{array}{c}           u_k \\ v_k \end{array}           \right]           e^{           i k j }           
\:\:\:\: , 
\label{appe.1.4}
\eneq
\noindent
with $- \frac{\pi}{2}           \leq k \leq \frac{\pi}{2}$, we eventually 
trade Eq.(\ref{appe.1.3}) for the analogous one in momentum 
space, given by 

\begin{eqnarray}
\epsilon u_k &=&  - ( J_e + J_o ) \cos ( k )             u_k  + i ( J_e - J_o )  \sin ( k ) v_k
\label{appe.1.5}\\
\epsilon v_k &=& -   i ( J_e - J_o )   \sin ( k ) u_k + ( J_e + J_o ) \cos ( k )          v_k  \nonumber 
\:\:\:\: . 
\end{eqnarray}
\noindent
The allowed values of $\epsilon$ are, therefore 

\beq
\pm \epsilon_k = 
 \pm   \sqrt{           (J_e + J_o )^2  \cos^2 ( k ) + (J_e - J_o)^2 \sin^2 ( k ) } 
\:\:\:\: , 
\label{appe.1.6}
\eneq
\noindent
with the corresponding wavefunctions  given by

\begin{eqnarray}
\psi_{ j , k , + } &=& c (-1)^j e^{ (-1)^j \frac{ i \varphi_k}{2}} e^{                i k j }                \nonumber \\
\psi_{ j , k , -} &=& c e^{ (-1)^j \frac{ i \varphi_k}{2}} e^{                i k j }                
\:\:\:\: . 
\label{appe.1.7}
\end{eqnarray}
\noindent
In Eq.(\ref{appe.1.7}) $c$ is a normalization constant and 

\begin{eqnarray}
\cos ( \varphi_k ) &=& \frac{                (J_e + J_o ) \cos ( k ) }{                \epsilon_k}                \nonumber \\
\sin ( \varphi_k ) &=& \frac{                (J_e - J_o ) \sin ( k ) }{\epsilon_k }                
\:\:\:\: , 
\label{appe.1.7}
\end{eqnarray}
\noindent
with  $- \frac{\pi}{2}                \leq k \leq \frac{\pi}{2}$.  
Due to the apparent degeneracy of the energy eigenvalues under  $k \to -k$, we may combine degenerate solutions 
with  opposite values of $k$, so to construct wavefunctions satisfying the open boundary conditions. These
are given by 

\begin{eqnarray}
\psi_{ j , k , + } &=&   (-1)^j \left\{    \alpha_{ k , + }   e^{ (-1)^j \frac{ i \varphi_k}{2}} e^{                i k j }                + \beta_{k , + }   
 e^{-  (-1)^j \frac{ i \varphi_k}{2}} e^{-                 i k j }                   \right\}    \nonumber \\
\psi_{ j , k , - } &=& \left\{    \alpha_{ k , - }  e^{  (-1)^j \frac{ i \varphi_k}{2}} e^{                i k j }                + \beta_{k , - }   
 e^{ - (-1)^j \frac{ i \varphi_k}{2}} e^{-                 i k j }                   \right\} 
\:\:\:\: , \nonumber \\
\label{appe.a.1}
\end{eqnarray}
\noindent
with $\alpha_{k , \pm}     , \beta_{k , \pm}$ constants to be determined by imposing  open boundary conditions,
that is   

 \beq
\psi_{ j = 0 , k , \pm } = \psi_{ j = L+1 , k , \pm }
\:\:\:\: . 
\label{appe.a.2}
\eneq
\noindent
Eqs.(\ref{appe.a.1},\ref{appe.a.2}) imply  the secular equation for $k$ given by

\beq
\sin [ k ( L + 1 ) - \varphi_k ] = 0 
\:\:\:\: . 
\label{appe.a.3}
\eneq
\noindent
Once $k$ satisfies Eq.(\ref{appe.a.3}), the solutions in Eq.(\ref{appe.a.1}) 
take the form 

\begin{eqnarray}
\psi_{ j , k , + } &=&  c  (-1)^j \left\{    e^{[ (-1)^j -1]  \frac{ i \varphi_k}{2}} e^{                i k j }                -  
 e^{- [ (-1)^j - 1]  \frac{ i \varphi_k}{2}} e^{-                 i k j }                   \right\}    \nonumber \\
\psi_{ j , k , - } &=& c \left\{   e^{ [  (-1)^j - 1]  \frac{ i \varphi_k}{2}} e^{                i k j }                - 
 e^{ - [ (-1)^j - 1]  \frac{ i \varphi_k}{2}} e^{-                 i k j }                   \right\} 
\:\:\:\: , \nonumber \\
\label{appe.a.4}
\end{eqnarray}
\noindent
with $c$ being an over-all normalization constant. 
In addition to the solutions discussed above, it is also possible to recover solutions with energy $ | \epsilon | < 2 | J_e - J_o |$. To 
derive them,  we set $k = \frac{\pi}{2}  - i q $. Accordingly, we obtain for the dispersion relation 

\beq
\epsilon_q^2 = (J_e - J_o )^2 \cosh^2 ( q ) - (J_e + J_o )^2  \sinh^2 ( q ) 
\;\;\;\; .
\label{appe.1.14}
\eneq
\noindent
Eqs.(\ref{appe.1.5}) now become 

\begin{eqnarray}
 \epsilon  u_q &=&  -  i ( J_e + J_o )  \sinh ( q ) u_q + i  (J_e - J_o )  \cosh ( q ) v_q \nonumber \\
 \epsilon  v_q &=&  -   i  (J_e - J_o )  \cosh ( q ) u_q + i ( J_e + J_o ) \sinh ( q ) u_q 
\;\;\;\; . \nonumber \\
\label{appe.1.16}
\end{eqnarray}
\noindent
Setting  

\beq
 e^{    -  i \varphi_\xi} = \left\{ \left( \frac{ J_e + J_o}{J_e - J_o} \right)  
     \tanh ( q ) - \frac{i \epsilon_q}{ ( J_e - J_o ) \cosh ( q ) }                \right\}  
 \:\:\:\: ,
 \label{novel.1}
 \eneq
\noindent
we eventually find   the real-space wavefunctions   in the form

\begin{eqnarray}
  \psi_{j,0,+}  &=&     \Biggl\{  \alpha_{+ , 0}  [ i^j e^{  \frac{ i \varphi_\xi}{2}} +  i^{-j} e^{ \frac{ -   i \varphi_\xi}{2}} ] e^{ q j }  
          \nonumber \\  &+&\beta_{+ , 0 }   [   i^j e^{ - \frac{ i \varphi_\xi}{2} } -   i^{-j} e^{   \frac{ i \varphi_\xi}{2} }  ] 
     e^{-            q  j   }  \Biggr\} \nonumber \\
      \psi_{j,0,-}  &=&  
    \Biggl\{ \alpha_{- , 0}      [   i^j    e^{-  \frac{ i  \varphi_\xi}{2} }+  i^{-j} e^{    \frac{ i \varphi_\xi}{2} }  ] 
e^{ q j }   \nonumber \\      &+& \beta_{- , 0}   [ - i^j   e^{  \frac{ i \varphi_\xi}{2}}+   i^{-j} e^{ - \frac{ i \varphi_\xi}{2}} ] 
     e^{-  q  j   }  \Biggr\}
\:\: .
\label{appe.a.x.1} 
\end{eqnarray}
\noindent
The allowed value of $q$ is determined again by the secular equation, which is now given by  

\beq
\tanh[ q ( L + 1 ) ] -  \cos ( \varphi_\xi ) = 0
\:\:\:\: .
\label{appe.addi.x1}
\eneq
\noindent
Eq.(\ref{appe.addi.x1})  takes a real solution for $q$  only   if  $J_e - J_o >  0$, 
which is the necessary condition to recover the topological phase. The wavefunctions for the ingap modes obeying 
open boundary conditions are given by

 \begin{eqnarray}
  \psi_{j,0,+}  &=&   c  \Biggl\{  \sin \left( \frac{\varphi_\xi}{2}    \right)  [ i^j e^{   \frac{ i \varphi_\xi}{2}} +     i^{-j} e^{- \frac{   i \varphi_\xi}{2}} ] e^{ q j }  
            \nonumber \\&+&
             i  \cos  \left( \frac{\varphi_\xi}{2}    \right)  [   i^j e^{ - \frac{ i  \varphi_\xi}{2} } -   i^{-j} e^{  \frac{ i \varphi_\xi}{2} }  ] 
     e^{-            q  j   }  \Biggr\} \nonumber \\
      \psi_{j,0,-}  &=&  
   c  \Biggl\{  \sin \left( \frac{\varphi_\xi}{2}    \right)    [   i^j   e^{-  \frac{ i  \varphi_\xi}{2} }+  i^{-j} e^{    \frac{ i \varphi_\xi}{2} }  ] 
e^{ q j }        \nonumber \\&+& i  \cos \left( \frac{\varphi_\xi}{2}    \right)   [-  i^j   e^{  \frac{ i \varphi_\xi}{2}} +   i^{-j} e^{  - \frac{ i \varphi_\xi}{2}} ] 
     e^{-  q  j   }  \Biggr\}
\:\: .
\label{appe.a.x.2} 
\end{eqnarray}
\noindent
The results presented in this appendix are what we have used in the main paper to compute quantities concerning 
the SSH chain at equilibrium.

\subsection{Scattering amplitudes in the presence of an impurity}
\label{scaam}

We now derive the scattering amplitudes in a nonuniform chain, in the presence of either a bond-, or of a dimer-impurity.
Let us begin by considering a single bond impurity: this is defined by modifying the strength 
of a single  odd bond  from $J_o \to J_o + \delta J$. Focusing, for the time being, on the 
solution  $\psi_{j , k , +}$  and without considering specific boundary conditions,  we 
construct  a scattering solution with momentum $k$ 
incoming from the left-hand side in the presence of a bond  impurity. Assuming that the impurity  is located across sites $j=1$ and $j=2$
and correspondingly alleging for a possible discontinuity of  $\psi_{j , k , +}$ across the impurity,   we set

\beq
\psi_{j , k , + }          = \Biggl\{          \begin{array}{l}          \psi_{j , k , +}^< \;\; , \; ( {\rm for}          \:  j \leq 1) \\
          \psi_{j , k , +}^>\;\; , \; ( {\rm for}          \:  j \geq 2) 
\end{array}
\:\:\:\:,
\label{mose.1}
\eneq
\noindent
with 
$\psi_{j,k,+}^{<(>)} = c \{ u_{j,k,+}^{<(>)} + (-1)^j  v_{j,k,+}^{<(>)} \}$, $c$ being a 
normalization constant and 

\begin{eqnarray}
u_{j , k , +}^< &=& (-1)^j i \sin \left( \frac{\varphi_k}{2} \right) \{ e^{ i k j} - r_k e^{ - i k j}          \}          \nonumber \\
u_{j , k , +}^> &=& (-1)^j i \sin \left( \frac{\varphi_k}{2} \right) t_k    e^{ i k j}            \nonumber \\
v_{j , k , +}^< &=& (-1)^j  \cos \left( \frac{\varphi_k}{2} \right) \{ e^{ i k j} +  r_k e^{ - i k j}          \}          \nonumber \\
v_{j , k , +}^> &=& (-1)^j \cos \left( \frac{\varphi_k}{2} \right) t_k    e^{ i k j}  
\:\:\:\: , 
\label{mose.2}
\end{eqnarray}
\noindent
with $r_k , t_k$ being the scattering amplitudes.
 The interface conditions across the impurities are given by  

\begin{eqnarray}
&& - (J_o + \delta J ) \psi_{2,k,+}^> + J_o \psi_{2,k,+}^<  = 0 \nonumber \\
&& J_o \psi_{1,k,+}^> - (J_o + \delta J ) \psi_{1,k,+}^<  = 0 
\:\:\:\: . 
\label{mose.3}
\end{eqnarray}
\noindent
Inserting Eqs.(\ref{mose.2}) into Eqs.(\ref{mose.3}), we recover 
the system for the scattering amplitudes in the form 
\begin{eqnarray}
&& - ( J_o + \delta J ) e^{          2 i k + \frac{ i \varphi_k}{2}} t_k + J_o \{          e^{          2 i k + \frac{ i \varphi_k}{2}} + r_k 
e^{          - 2 i k -  \frac{ i \varphi_k}{2}} \} = 0 \nonumber \\
&& J_o e^{           i k -  \frac{ i \varphi_k}{2}} t_k - ( J_o + \delta J )  \{          e^{            i k -  \frac{ i \varphi_k}{2}} + r_k 
e^{          -   i k+   \frac{ i \varphi_k}{2}} \} = 0
\:\:\:\: . \nonumber \\
\label{mose.5}
\end{eqnarray}
\noindent
Finally, from Eq.(\ref{mose.5}) we obtain

\beq
r_k = - \frac{ e^{ 4 i k + i \varphi_k} \delta J ( 2J_o + \delta J ) }{ - J_o^2 + e^{ 2 i ( k + \varphi_k ) } ( J_o + \delta J )^2 } 
\;\;\;\; .
\label{mose.7}
\eneq
\noindent
In terms of the parameter $W$ introduced in Eq.(\ref{model.3}) and setting $J_o = 1$, we 
reexpress Eq.(\ref{mose.7}) as

\beq
r_k =  \frac{ e^{ 4 i k + i \varphi_k} W ( 2-W  ) }{ - 1 + e^{ 2 i ( k + \varphi_k ) } ( 1-W )^2 } 
\;\;\;\; . 
\label{mose.7bis}
\eneq
\noindent
Eq.(\ref{mose.7bis}) is what we use in section \ref{ines_bond}         to discuss the impurity-induced 
localization in the presence of bond disorder. 

In the case in which there is a dimer-impurity   at sites $j=1$ and 
$j=2$  into an  otherwise uniform SSH chain, 
we see that the interface conditions in Eq.(\ref{mose.3})  are   substituted by the conditions 

 \begin{eqnarray}
&& -W \psi_{1,k,+}^< + J_o ( \psi_{2,k,+}^> -  \psi_{2,k,+}^< )  = 0 \nonumber \\
&& J_o (  \psi_{k,1,+}^< - \psi_{k,1,+}^> )- W \psi_{2,k,+}^> = 0 
\;\;\;\; . 
\label{mose.4}
\end{eqnarray}
\noindent
  with $W$ denoting the corresponding potential strength, 

Choosing a positive-energy solution as in Eqs.(\ref{mose.1},\ref{mose.2}), 
from  Eqs.(\ref{mose.4}) we obtain the system of equations for $r_k$ and $t_k$ in 
the form 

\begin{eqnarray}
&&- W  \{          e^{            i k -  \frac{ i \varphi_k}{2}} + r_k 
e^{          -   i k+   \frac{ i \varphi_k}{2}} \} \nonumber \\
&+& J_o\{           e^{          2 i k + \frac{ i \varphi_k}{2}} t_k 
-   e^{          2 i k + \frac{ i \varphi_k}{2}} -  r_k 
e^{          - 2 i k -  \frac{ i \varphi_k}{2}} \}  = 0 \nonumber \\
&&  J_o \{          e^{            i k -  \frac{ i \varphi_k}{2}} + r_k 
e^{          -   i k+   \frac{ i \varphi_k}{2}} - e^{           i k -  \frac{ i \varphi_k}{2}} t_k  \} \nonumber \\
&-& W   e^{          2 i k + \frac{ i \varphi_k}{2}} t_k = 0 
\:\:\:\: . 
\label{mose.6}
\end{eqnarray}
\noindent
Solving for $r_k$, we now obtain 
 
\beq
r_k = - \frac{ e^{ 4 i k + i \varphi_k} W (  W + 2 J_o  \cos ( k + \varphi_k) )}{ 
J_o^2 + 2 e^{ i ( k + \varphi_k ) } J_o W + e^{ 2 i ( k + \varphi_k ) } ( - J_o^2 + W^2 ) }  
\;\;\;\; . 
\label{mose.8}
\eneq
\noindent
Eq.(\ref{mose.8}) is what we have used in section \ref{ines_dimer} to discuss the
localization effects of dimer disorder in the SSH chain and, in particular, to infer
under which conditions we recover the emergence of a mobility edge in the disordered system.

\bibliography{biblio_ssh}

\end{document}